\documentclass[useAMS,usenatbib,usegraphicx]{mn2e}

\title[The nature of the red disk-like galaxies at high redshift]
  {The nature of the red disk-like galaxies at high redshift:\\
dust attenuation and intrinsically red stellar populations}
\author[D. Pierini et al.]
  {D.~Pierini,$^{1}$ 
  C.~Maraston,$^{2}$ 
  K.~D.~Gordon,$^{3}$ 
  and A.~N.~Witt$^{4}$ \thanks{E-mail: dpierini@mpe.mpg.de, maraston@astro.ox.ac.uk, kgordon@as.arizona.edu, awitt@dusty.astro.utoledo.edu}
\\
$^{1}$Max-Planck-Institut f\"ur extraterrestrische Physik, Giessenbachstrasse, Garching b. M\"unchen, D-85748, Germany\\
$^{2}$University of Oxford, Denys Wilkinson Building, Keble Road, Oxford, OX1 3RH, UK\\
$^{3}$Steward Observatory, University of Arizona, Tucson, AZ 85721, USA\\
$^{4}$Ritter Astrophysical Research Centre, The University of Toledo, Toledo, OH 43606, USA}
\begin{document}

\date{Accepted ... . Received ... ; in original form ...}

\pagerange{\pageref{firstpage}--\pageref{lastpage}} \pubyear{...}

\maketitle

\label{firstpage}

\begin{abstract}
We investigate which conditions of dust attenuation and stellar populations
allow models of dusty, continuously star-forming, bulge-less disk galaxies
at $0.8 \la z \la 3.2$ to meet the different colour selection criteria
of high-$z$ ``red'' galaxies (e.g. $\rm R_c - K > 5.3$, $\rm I_c - K > 4$,
$\rm J - K > 2.3$).
As a main novelty, we use stellar population models that include
the thermally pulsating Asymptotic Giant Branch (TP-AGB) phase
of stellar evolution.
The star formation rate of the models declines exponentially
as a function of time, the e-folding time being longer than 3 Gyr.
In addition, we use calculations of radiative transfer of the stellar
and scattered radiation through different dusty interstellar media
in order to explore the wide parameter space of dust attenuation.
We find that synthetic disks can exhibit red optical/near-infrared colours
because of reddening by dust, but only if they have been forming stars
for at least $\sim$ 1 Gyr.
Extremely few models barely exhibit $\rm R_c - K > 5.3$,
if the inclination $i = 90$\degr
and if the opacity $2 \times \tau_{\rmn V} \ga 6$.
Hence, $\rm R_c - K$-selected galaxies at $1 \la z \la 2$
most probably are either systems with an old, passively evolving bulge
or starbursts.
Synthetic disks at $1 \la z \la 2$ exhibit $\rm 4 < I_c - K < 4.8$,
if they are seen edge on (i.e. at $i \sim 90$\degr)
and if $2 \times \tau_{\rmn V} \ga 0.5$.
This explains the large fraction of observed, edge-on disk-like galaxies
with $\rm K_s < 19.5$ and $\rm F814W - K_s \ga 4$.
Finally, models with $2 \la z \la 3.2$ exhibit $\rm 2.3 < J - K < 3$,
with no bias towards $i \sim 90$\degr and for a large range in opacity
(e.g. $2 \times \tau_{\rmn V} > 1$ for $i \sim 70$\degr).
In conclusion, red disk-like galaxies at $0.8 \la z \la 3.2$
may not necessarily be dustier than nearby disk galaxies
(with $0.5 \la 2 \times \tau_{\rmn V} \la 2$) and/or much older
than $\sim$ 1 Gyr.
This result is due both to a realistic description of dust attenuation
and to the emission contribution by TP-AGB stars,
with ages of 0.2 to 1--2 Gyr and intrinsically red colours.
\end{abstract}


\section{Introduction}

The standard hierarchical merging scenario of structure formation and evolution
(White \& Rees 1978; White \& Frenk 1991) is conducive to an initial formation
of disks, in view of the large predicted collapse factors
(Fall \& Efstathiou 1980).
Model stellar disks are rotationally supported and have exponential,
radial surface-brightness profiles, as observed (Mo, Mao, \& White 1998).
A bulge$+$disk system forms in a two-step process: merging two equal-mass disk
progenitors produces a central spheroidal component, the bulge, in the same way
as an elliptical galaxy forms (see Barnes \& Hernquist 1992);
then gas from the surrounding dark halo cools and settles to form a new disk.
Since disks of bulge$+$disk systems in the local Universe are a product
of accretion of gas over a Hubble time, late-type spirals must form
their bulges at higher redshifts than earlier types (Kauffmann 1996).
In simulations of the evolution of the cosmic star-formation rate (SFR)
in cold dark matter (CDM) universes, half of all stars form at $z \ga 2.2$
(Hernquist \& Springel 2003).
Therefore, {\em disk galaxies} (with prolonged star-formation activity),
{\em starburst galaxies} (with high intensity, short-lived star-formation
activity), and {\em starbursts triggered by merging} constitute
the main star factories at $z \ga 2$ for hierarchical merging cosmologies.

Observed optical/near-infrared (IR) colours like $\rm R - K$, or $\rm I - K$,
and $\rm J - K$ are used to select and classify statistically
high-$z$ ``red'' objects.
In particular, red objects at $1 \la z \la 2$ are expected to include
systems dominated by old (i.e. with age $>$ 1--2 Gyr), passively evolving
stellar populations, with a spectral energy distribution (SED) exhibiting
a well developed $\rm 4000~\AA$-break (rest frame), usually identified
as early-type galaxies, or, conversely, dusty gas-rich systems
with a continuous star-formation activity, i.e. disk galaxies
(Pozzetti \& Mannucci 2000).
In addition, they are expected to include dusty gas-rich systems
with an impulsive, very intense star-formation activity (i.e. starbursts),
and the early, dusty post-starburst phases of galaxy formation
that started at $1 < z < 3$ (Pierini et al. 2004a).

As a matter of fact, the observed {\em extremely red galaxies}
(with e.g. $\rm I - K > 4$ or $\rm R - K > 5.3$) are very heterogeneous
(Smail et al. 2002; Yan \& Thompson 2003; Cimatti et al. 2003;
Stevens et al. 2003; Gilbank et al. 2003; Moustakas et al. 2004;
see McCarthy 2004 for a review).
Morphological classifications using surface brightness profile-fitting
(Cimatti et al. 2003; Gilbank et al. 2003) show that
the fraction of bright (i.e. with $\rm K < 20$) $\rm I - K$-selected galaxies
exhibiting clear disk components is about $35 \pm 11$ per cent.
They also confirm that a further 30--37 per cent
is made of spheroidal or compact systems (see Yan \& Thompson 2003),
while at least 15 per cent is made of disturbed or irregular objects.
No link between morphological classification and spectroscopic-redshift
distribution ($0.7 < z < 1.5$) exists for bright, extremely red galaxies
(Cimatti et al. 2002; Yan, Thompson, \& Soifer 2004).

For all these reasons, it is puzzling that
extremely red optical/near-IR colours are never achieved
by models of an evolving, dusty Sb-type galaxy at $z \la 3$,
according to V\"ais\"anen \& Johansson (2004).

Interestingly, 40 per cent of the Yan \& Thompson (2003) disky systems
(i.e. either disk-dominated or bulge-less) with $\rm K_s < 19.5$
and $\rm F814W - K_s \ga 4$ are viewed almost edge on
(i.e. at an inclination of $\sim$ 90\degr).
Their extremely red colours are tentatively attributed
to the attenuation of stellar light by dust distributed in their disks.
According to Yan \& Thompson, several of these edge-on disky systems
possibly lie at $z < 1$, given that their large apparent sizes
(several arcseconds) would lead to unusually large physical sizes
(tens of kiloparsecs) for a hypothetical redshift $\ga 1$.
However, bright, $\rm F814W - K_s$-selected disky galaxies do exist
at $z \sim 1$ (spectroscopic, see Yan et al. 2004).

Furthermore, Labb\'e et al. (2003) find 6 galaxies at photometric redshifts
between 1.4 and 3, with $\rm K_s$-band surface brightness profiles
consistent with an exponential law over 2--3 effective radii,
and with face-on (i.e. viewed at an inclination of 0\degr) effective radii
comparable to that of the Milky Way (see also Conselice et al. 2004).
All 6 large disk-like galaxies have a ``red'' nucleus,
while their optical/near-IR colours become bluer in the outer parts.
Labb\'e et al. (2003) can not provide a definitive interpretation
of these observables, but suggest several ones: more dust, higher age,
emission-line contamination, or a combination of effects,
in the galaxy centres.
In particular, one large disk-like galaxy at $z \sim 3$ (photometric)
has $\rm K_s = 20.53$, $\rm F814W - K_s = 3.9$, and $\rm J_s - K_s = 2.6$
(F\"orster Schreiber et al. 2004).
Interestingly, Stockton, Canalizo, \& Maihara (2004) have discovered
another disk-like galaxy at $z \sim 2.5$ (photometric)
with $\rm J - K^{\prime} \sim 3.4$ and an inclination of about $70$\degr.

Objects with $\rm J_s - K_s > 2.3$ are expected to be at $z > 2$,
their colours being due to age and/or dust effects (Franx et al. 2003).
Thus, they have been called {\em distant red galaxies}
(van Dokkum et al. 2004).
Most of the distant red galaxies studied so far exhibit red near-IR colours
in the range 2.3--3.9 mag possibly as a result of a prominent Balmer break
and/or $\rm 4000~\AA$-break moving into the $\rm J_s$ band,
independent of their star-formation activity
(Franx et al. 2003; van Dokkum et al. 2004; F\"orster Schreiber et al. 2004).
They seem to be among the oldest (median age $\sim$ 1--2 Gyr) and most massive
(median stellar mass $\rm \sim 10^{11}~M_{\sun}$) galaxies
at $2 \la z \la 3.5$, with a median SFR $\sim$ 15--$\rm 150~M_{\sun}~yr^{-1}$
and a median rest-frame V-band attenuation $A_{\rmn V} \sim$ 0.9--2.5 mag,
according to the assumed star-formation history
(F\"orster Schreiber et al. 2004; see also Toft et al. 2005).

However, these authors assume that the empirical ``Calzetti law''
for nearby starbursts (Calzetti, Kinney, \& Storchi-Bergmann 1994;
Calzetti et al. 2000) describes dust attenuation for any high-$z$,
dusty stellar system.
This key assumption for the SED modeling is questionable.
In the local Universe, the characteristics of dust attenuation are different
between starbursts, normal star-forming (i.e. non starburst) disk galaxies
(Bell 2002; Gordon et al. 2004), and ultra-luminous infrared galaxies
(Goldader et al. 2002).
In conclusion, investigating the nature of distant, red disk-like galaxies
(hereafter referred to as DRdGs, with $\rm J - K > 2.3$) or of extremely red
disk-like galaxies (hereafter referred to as ERdGs, with $\rm I_c - K > 4$)
can shed new light on the formation and evolution of disk systems.

Therefore, it is fundamental to understand if the optical/near-IR colours
of DRdGs and ERdGs are due either to the luminosity-weighted age
of the stellar populations present in the system or to reddening
by internal dust.

To this purpose, we combine evolutionary synthesis models
of stellar populations (Maraston 1998, 2005) with models of radiative transfer
of the stellar and scattered radiation through the dusty interstellar medium
(ISM) of a disk system (Ferrara et al. 1999; Pierini et al. 2004b).

The stellar population evolutionary synthesis models used here include
the thermally pulsating Asymptotic Giant Branch (TP-AGB) phase
for intermediate-mass (2--5 $\rm M_{\sun}$) stars, at variance with all others
(Maraston 2005).
This phase is fundamental for computing correctly the SEDs of intermediate-age
(i.e. between 0.2 and 1--2 Gyr) stellar populations (see Maraston 1998, 2005).

The radiative transfer models (including multiple scattering)
for the disk geometry used here describe
how the attenuation function\footnote{The extinction curve describes
the combined absorption and out-of-the beam scattering properties
of a mixture of dust grains of given size distribution and chemical composition
in a screen geometry as a function of wavelength; the attenuation function
is the combination of the extinction curve with the geometry of a dusty
stellar system, in which a substantial fraction of the scattered light
is returned to the line of sight. We refer the reader to Calzetti (2001)
for a discussion in more detail.} changes in absolute value and shape
with total amount of dust and inclination of the disk, for different dust/stars
configurations and different structures of the dusty ISM.

\section{Characteristics of the models}

\subsection{Stellar populations}

We assume that stellar populations are formed
starting from an initial star-formation redshift $z_{\rmn f}$ equal to 10.
This value is between the two redshifts considered by Ciardi \& Madau (2003)
as markers of the epoch of reionization breakthrough
in their `` late reionization'' and `` early reionization'' scenarios.
We assume the so-called $\rm \Lambda$CDM ``concordance'' cosmological model
($\Omega_{\rmn m} = 0.3$, $\Omega_{{\rmn \Lambda}} = 0.7$) with Hubble constant
$H_0 = 70~\rm km~s^{-1}~Mpc^{-1}$, consistent with the main {\it WMAP} results
(Spergel et al. 2003), in order to link age (i.e. the time that has elapsed
since star formation started) and redshift.

Observed red disk-like galaxies span the redshift range 0.7--3,
so that we consider six representative redshifts for candidate DRdGs and ERdGs,
i.e. $z = 0.83$, 1.07, 1.39, 1.85, 2.59, and 3.19, that correspond
to maximum ages of the stellar populations equal to 6, 5, 4, 3, 2, and 1.5 Gyr,
respectively, for $z_{\rmn f} = 10$.
We compute additional models, where the stellar populations have a maximum age
of 0.6 Gyr for any of the previous 6 redshifts, with the intent of constraining
the lowest value of the parameter $z_{\rmn f}$ that is necessary to reproduce
the observed optical/near-IR colours of DRdGs and ERdGs.

We model the intrinsic (i.e. non attenuated by internal dust) SED
of a dusty, bulge-less disk galaxy as a composite stellar population
with solar metallicity ($\rm Z_{\sun}$) and a Salpeter (1955) initial mass
function (IMF) between 0.1 and $\rm 120~M_{\sun}$.
Solar metallicities seem to be present in massive, star-forming galaxies
at $z \ga 2$ (Shapley et al. 2004; van Dokkum et al. 2004).
A ``normal'' IMF seems to be established when the ISM metallicity reaches
a value of $\rm \sim 10^{-4}~Z_{\sun}$ (Schneider et al. 2002).
Furthermore, the Salpeter IMF describes in a consistent way
different properties of the Universe at least up to $z \sim 1$ (Renzini 2004).

The intrinsic SEDs of the composite stellar populations are computed
with the evolutionary synthesis code of Maraston (1998, 2005).
The models include the contribution by the TP-AGB phase.
TP-AGB stars are cool giants exhibiting very red optical/near-IR colours
(e.g. Persson et al. 1983).
They provide about 40 per cent of the bolometric luminosity
and up to 80 per cent of the rest-frame K-band luminosity
of an intermediate-age simple stellar population (SSP) (Maraston 1998, 2005).
They are expected to play a significant role in the near-IR emission
of galaxies containing 1 Gyr-old stellar populations,
since the fuel consumption during the TP-AGB phase is maximum around this age
(Maraston 1998, 2005).
More generally, these stars must be included in any realistic modeling
of stellar populations, since intermediate-mass stars do unavoidably experience
the TP-AGB phase.

In the framework of this study, the effect of the TP-AGB stars
will matter when the R, I, J, and K bands (i.e. those used to select
high-$z$ red galaxies) map the rest-frame near-IR.
Therefore the lower redshift bins will be affected.
Since the TP-AGB phase is usually not included in population synthesis models
(see Maraston 2005 for examples), we have considered also some models
in which the TP-AGB phase is neglected.
This allows us to compare our results with others in the literature.
With respect to the latter, we will show in Sect. 3.2.2 that
the physically-motivated presence of TP-AGB stars softens the need
either of dominant, old, passively evolving stellar populations
or of large, arbitrary values of reddening by dust associated with high SFRs
to explain the observed red optical/near-IR colours of high-$z$ galaxies.

The time dependence of the SFR\footnote{The intrinsic SEDs
of the stellar population models are available
at www-astro.physics.ox.ac.ik/$\sim$maraston.}, is described by
$SFR(t) = \alpha \times e^{- \frac{t}{\tau_{\star}}}$ (Eq. 1),
where $\alpha$ is the peak SFR (in $\rm M_{\sun}~yr^{-1}$) and $\tau_{\star}$
is the SFR e-folding time, here set equal to 3, 4, 5, or 7 Gyr.
We will discuss qualitatively the effects due
to an alternative parameterization for the star-formation history,
i.e. (e.g. Moll\'a, Ferrini, \& D\'iaz 1997; Moll\'a, Ferrini, \& Gozzi 2000)
$SFR(t) = \alpha \times \frac{t}{\tau_{\rmn p}}~e^{- \frac{t}{\tau_{\rmn p}}}$
(Eq. 2), where $\alpha/e$ represents the peak SFR, $\tau_{\rmn p}$ is the time
when the SFR peaks, and the SFR increases linearly with time
for $t < \tau_{\rmn p}$, but decreases exponentially with time
for $t > \tau_{\rmn p}$\footnote{For reference, we give the values
of the peak SFR predicted by Eq. 1 and 2 for two significant examples.
The first case regards a bulge-less disk galaxy with a stellar mass
of $\rm 5 \times 10^{10}~M_{\sun}$ at $z = 0$ (like the Milky Way,
see Dehnen \& Binney 1998), that started forming stars at $z_{\rmn f} = 10$.
The peak SFR in Eq. 1 (i.e. $\alpha$) is equal to $(1 - R)^{-1} \times$ 16.9,
13, 10.8, or $\rm 8.5~M_{\sun}~yr^{-1}$, for $\tau_{\star}$ equal to 3, 4, 5,
or 7 Gyr, respectively, where $R$ is the fraction of mass
in a stellar generation that is returned to the ISM over the life time
of the population (Tinsley 1980).
For the same end-product, if Eq. 2 holds, the peak SFR (i.e. $\alpha/e$)
is equal to $(1 - R)^{-1} \times$ 6.6 or $\rm 5~M_{\sun}~yr^{-1}$
for $\tau_{\rmn p}$ equal to e.g. 3 or 5 Gyr, respectively.
The second case is represented by a bulge-less disk galaxy with a stellar mass
of $\rm 2 \times 10^{11}~M_{\sun}$ at $z \sim 2.4$
(see van Dokkum et al. 2004), that started forming stars at $z_{\rmn f} = 10$.
Now, the peak SFR in Eq. 1 is equal to $(1 - R)^{-1} \times$ 1.3, 1.2, 1.1,
or $\rm 1~\times~10^2~M_{\sun}~yr^{-1}$, for $\tau_{\star}$ equal to 3, 4, 5,
or 7 Gyr, respectively.  For the same end-product, the peak SFR in Eq. 2
is equal to $(1 - R)^{-1} \times$ 1.5 or $\rm 2~\times~10^2~M_{\sun}~yr^{-1}$
for $\tau_{\rmn p}$ equal to 3 or 5 Gyr, respectively.}.

Star-formation time scales as short as 3 Gyr may sound atypical,
if one wants to describe properties of local, ``quiescent'' star-forming
galaxies (see Kong et al. 2004).
However, the observations seem to point to a fast evolution of disk galaxies
at high $z$ down to $z \sim 1$ (Abraham \& Merrifield 2000;
Kajisawa \& Yamada 2001; Ravindranath et al. 2004), when the local age
of the Universe was about 5.7 Gyr.
Hence, a SFR e-folding time (Eq. 1) or a SFR characteristic time scale (Eq. 2)
significantly shorter than 5.7 Gyr offers a simple but reasonable
representation of the unknown star-formation history of a disk-like galaxy
observed at $z > 1$.

The bulge-less hypothesis is explored because of the observational evidences
reported in Sect. 1.
In addition, we note that hints come from recent evolutionary models
of star-forming, gas-rich disks (Immeli et al. 2004) that these systems
may remain bulge-less for a large part of the Hubble time,
if the energy dissipation of the cold cloud component is inefficient.
In this case, the global SFR has a time dependence that is broadly analogous
in shape to that given by Eq. 2, with a time scale longer than about 2 Gyr.
We may then expect that at least disk-galaxy models with $SFR(t)$
given by Eq. 2 and $\tau_{\rmn p}$ set equal to 3--5 remain bulge-less
within the redshift range under study.
We postulate that this is so also for models where $SFR(t)$ is given by Eq. 1
with $\tau_{\star} \rm \ga 3~Gyr$.

\subsection{Attenuation by internal dust}

We assume that the high-$z$, red disk-like galaxies
are axially symmetric systems, where the 3-D distribution either of stars
or of dust can be parameterized by a doubly exponential law.
The dust/stars configuration is preserved, whether the average disk size
(for fixed luminosity) changes as a function of redshift or not
(cf. Ferguson et al. 2004; Bouwens et al. 2004;
Ravindranath et al. 2004)\footnote{The exponential disk scale-length decreases
with $z$, for $z \la 1$, as $\sim (1 + z)^{-1}$ (for given rotation speed)
in hierarchical models for galaxy formation (Mao, Mo, \& White 1998).}.
The dust/stars configuration and the properties of the dusty ISM
of high-$z$ galaxies are not well constrained by existing observations.
Therefore, we adopt two different sets of Monte Carlo calculations
of radiative transfer for the disk geometry, that embrace
most of the parameter space defined by dust/stars configuration,
structure of the dusty ISM, and dust type, that is either explored
in the literature (see Bianchi 2004 for a recent review)
or known from observations of normal star-forming disk galaxies
in the local Universe (White, Keel, \& Conselice 2000; Keel \& White 2001;
Dalcanton, Yoachim, \& Bernstein 2004; Whittet 2003 and references therein).

The first set of simulations is taken from Pierini et al. (2004b);
it illustrates results from radiative transfer models for the bulge and disk
components of a late-type galaxy, based on the DIRTY code (Gordon et al. 2001).
These models have been applied successfully to interpret
multiwavelength photometry of edge-on late-type galaxies in the local Universe
(Kuchinski et al. 1998).
Here we summarize their relevant features.
The radial distribution is the same for stars and dust, while the scale height
of the stellar component decreases with decreasing wavelength
of the stellar emission.
The young stars responsible of the bulk UV emission are the most embedded
in the large-scale dust distribution, that can be considered analogous
to a geometrically-thin dust lane.
As for the structure of the dusty ISM, this can be either homogeneous
(i.e. diffuse) or two-phase, clumpy (i.e. the ISM contains diffuse gas
plus molecular clouds).
Dust clumps are distributed stochastically with respect to each other
and to the bluest stars; the most opaque ones are located in the inner disk.
The physical properties of the dust grains are assumed to be the same
as those in the diffuse ISM of the Milky Way (from Witt \& Gordon 2000).
The opacity\footnote{Hereafter, the term opacity refers
to the face-on extinction optical-depth through the centre of the dusty disk
at the V band. The opacity gives the total amount of dust
once the dust distribution and the disk sizes are fixed (see e.g. Ferrara
et al. 1999).} of the disk is equal to $2 \times \tau_{\rmn V}$,
with $\tau_{\rmn V}$ equal to 0.25, 0.5, 1, 2, 4, and 8.
For reference, the opacity of local disks ranges between 0.5 and 2
(Kuchinski et al. 1998).
Finally, only three values for the inclination $i$ of the disk are explored
here (0, 70, and 90 degrees), since reddening changes slowly with $i$,
for $i < 70$\degr (e.g. Pierini et al. 2004b).

The second set of simulations is taken from Ferrara et al. (1999).
There, the scale height of the stellar distribution does not depend
on the wavelength of the tracing photons, and both the scale height
and the scale length of the dust distribution are allowed to be larger
than those of the stellar distribution.
As these authors discuss, the dust-to-stars scale-height ratio
has a larger effect on the attenuation function than the dust-to-stars
scale-length ratio.
Therefore, we consider three dust/stars configurations, where dust and stars
have the same scale length, and the dust-to-stars scale-height ratio
is equal to 0.4 (corresponding to a geometrically-thin dust lane), 1, or 2.5.
Ferrara et al. consider only the presence of a dusty, diffuse (or homogeneous)
ISM, but with two choices for the mixture of dust grains, namely like
in the Milky Way (MW) or in the Small Magellanic Cloud (SMC),
as given by Gordon, Calzetti, \& Witt (1997).
The two dust types produce very different extinction curves
at rest-frame wavelengths shorter than $\rm 2500~\AA$, which impacts
on the observed $\rm I_c$ and $\rm R_c$ magnitudes of objects with redshift
down to 2.5 and 1.7, respectively.
The continuum of observed dust extinction curves is possibly caused
by the environmental stresses of nearby star-formation activity
(Gordon et al. 2003).
Finally, we consider $\tau_{\rmn V}$ equal to 0.05, 0.25, 0.5, 1, 2.5, and 5,
and set $i$ equal to 9.3, 70, and 90 degrees.

Both sets of Monte Carlo computations of radiative transfer assume that
the dustiest region of a disk is its centre.
Hence, they consider the hypothesis that the nuclear, redder SEDs
of the 6 high-$z$, large disk-like galaxies discovered by Labb\'e et al. (2003)
are due to a larger dust attenuation.
Conversely, these calculations do not take into account the existence
of a radial gradient in the distribution of the stellar populations,
a second possibility suggested by Labb\'e et al. as a cause
for the colour gradients existing between the inner and outer regions
of their sample galaxies.

\subsection{Dimming due to foreground media}

We do not apply corrections for dimming due either
to continuum and line absorption by atomic hydrogen (H{\small I}),
or to line absorption by metals, or to continuum absorption by dust,
where the H{\small I} gas, metals, and dust are distributed in foreground media
(clumpy or diffuse) of galactic or intergalactic nature.
Hereafter we explain our reasons.

For $z \sim 3.2$ (i.e. the maximum redshift considered for our models),
the H{\small I} Ly$\rm \alpha$ transition at $\rm 1216~\AA$ (rest frame)
is red-shifted to $\rm \sim 5107~\AA$, thus shortward of the wavelength domain
probed by the $\rm R_c$ broad-band filter.
Hence, the effects of the stochastic attenuation produced
by intervening H{\small I} gas (Madau 1995) do not affect at all
the computation of the observed $\rm R_c$ magnitude and, a fortiori,
of the other observed magnitudes considered in this study (see Tab. 2).

On the other hand, metal-line absorption associated with foreground systems
(e.g. the damped Ly$\rm \alpha$ systems, see Dessauges-Zavadsky et al. 2004)
is expected to affect in a negligible way
the observed optical/near-IR magnitudes and colours of galaxies
at $0.8 \la z \la 3.2$.
The cumulative effect on the optical/near-IR broad-band photometry of galaxies
at these redshifts due to absorption by dust associated
with individual damped Ly$\rm \alpha$ systems, stochastically distributed
along the line of sight, is also expected to be very small
for the dust-to-gas ratios typically inferred in these systems
(see Charlot \& Fall 1991; Haardt \& Madau 1996).

However, continuum absorption by dust distributed
in the diffuse intergalactic medium (IGM) can be a source of concern
especially for the observed $\rm R_c$ and $\rm I_c$ magnitudes
and, thus, for the observed $\rm R_c - K$ and $\rm I_c - K$ colours
of objects at $0.8 \la z \la 3.2$.
In particular, this is true if the amount of IGM dust is close
to the maximum value estimated by Inoue \& Kamaya (2004), which is constrained
by extinction and reddening of distant Type Ia supernovae
and by the thermal history of the IGM affected by dust photoelectric heating.
The physics of the ejection of dust from the ISM of a galaxy to the IGM
is complex (Aguirre 1999; Bianchi \& Ferrara 2005 and references therein)
and opens quite different possibilities as for the IGM extinction law
(see Aguirre 1999; Inoue \& Kamaya 2004; Bianchi \& Ferrara 2005).
For instance, if only grains with sizes larger than $\rm \sim 0.1~\mu m$
populate the IGM with an abundance large enough to account
for the observed Type Ia supernova dimming at $\rm z \sim 0.5$ (Aguirre 1999),
we expect the observed $\rm R_c - K$ and $\rm I_c - K$ colours
computed for our models with $2 \la z \la 3.2$ to be underestimated
by up to a few tenths of a magnitude.
This underestimate would be important, if true.
However, the preferred interpretation for the dimming of Type Ia supernovae
is the cosmological origin (see de Bernardis et al. 2000; Spergel et al. 2003).
Furthermore, despite the present knowledge of the dusty IGM is poor,
a dynamical model for the metal enrichment of the IGM at $z = 3$
through dust sputtering (Bianchi \& Ferrara 2005) suggests that
extinction and gas photoelectric heating effects due to IGM dust grains
are well below current detection limits.

In conclusion, the computed $\rm R_c$, $\rm I_c$, J, and K magnitudes,
and $\rm R_c - K$, $\rm I_c - K$, and $\rm J - K$ colours reflect directly
the properties assumed either for the stellar populations or for the dusty ISM
of the high-$z$ disks.

\subsection{The final grid of models}

Table 1 lists the characteristics of selected models of dusty,
star-forming disk galaxies that will be discussed hereafter.
\begin{description}
 \item
 Column 1 shows the model index, where a different capital letter
 refers to the value assumed for the SFR e-folding time $\tau_{\star}$,
 and a different figure indicates either the dust/stars configuration
 or the structure of the dusty ISM of choice.
 \item
 Column 2 shows the value assumed for $\tau_{\star}$ (in Gyr).
 \item
 Column 3 indicates the dust/stars configuration of choice.
 ``Standard'' refers to the simulations by Pierini et al. (2004b),
 where the youngest stars are the most embedded in a narrow dust lane.
 The other acronyms refer to the simulations by Ferrara et al. (1999),
 where dust and stars are distributed with the same scale length
 but with a dust-to-stars scale-height ratio equal to
 0.4 (``S01\_ME04'' and ``S01\_SE04''), 1 (``S01\_ME10'' and ``S01\_SE10''),
 or 2.5 (``S01\_ME25'' and ``S01\_SE25'').
 \item
 Column 4 indicates whether the structure of the dusty ISM is two-phase clumpy
 (``c'') or homogeneous (``h'').
 \item
 Column 5 indicates whether the dust is either of MW type (``MW'')
 or of SMC type (``SMC'').
 \item
 Column 6 shows the adopted radiative transfer model.
\end{description}

Each set of models contains 108 realizations, computed for 6 values
of $\tau_{\rmn V}$, 3 values of $i$, and 6 values of $z$, for a total
of 3348 simulated high-$z$, dusty, continuously star-forming disks.
We recall that the 6 redshifts correspond either to 6 different maximum ages
if $z_{\rmn f} = 10$ (models A1 through B8), or to a fixed maximum age
(i.e. 0.6 Gyr), $z_{\rmn f}$ being a variable (models C1).

For these 6 redshifts, Table 2 shows how the rest-frame wavelength domain
is mapped by the central wavelengths of the $\rm R_c$, $\rm I_c$, J, and K
broad-band filters (Cousins 1978; Bessel \& Brett 1988).

\begin{table}
 \centering
 \begin{minipage}{84mm}
  \caption{Model characteristics}
  \begin{tabular}{@{}cclccc@{}}
  \hline \hline
   Den. & $\tau_{\star}$ & dusty disk & ISM & dust type & ref.$^{a}$\\
 \hline \hline
A1 & 5.0 & standard & c & MW & 1 \\
A2 & 5.0 & standard & h & MW & 1 \\
B1 & 3.0 & standard & c & MW & 1 \\
B2 & 3.0 & standard & h & MW & 1 \\
B3 & 3.0 & S01\_ME04 & h & MW & 2 \\
B4 & 3.0 & S01\_ME10 & h & MW & 2 \\
B5 & 3.0 & S01\_ME25 & h & MW & 2 \\
B6 & 3.0 & S01\_SE04 & h & SMC & 2 \\
B7 & 3.0 & S01\_SE10 & h & SMC & 2 \\
B8 & 3.0 & S01\_SE25 & h & SMC & 2 \\
C1 & 5.0 & standard & c & MW & 1 \\
  \hline
  \end{tabular}
 \medskip
 \\
 $^{a}$ 1: Pierini et al. (2004b); 2: Ferrara et al. (1999)
 \end{minipage}
\end{table}

\begin{table}
 \centering
 \begin{minipage}{84mm}
  \caption{Observed frame vs. rest frame}
  \begin{tabular}{@{}ccccc@{}}
  \hline \hline
   z & $\rm R_c$ & $\rm I_c$ & J & K \\
 \hline \hline
0.83 & 3535$^{a}$ & 4298$^{a}$ & 6776$^{a}$ & 12076$^{a}$ \\
1.07 & 3125 & 3799 & 5904 & 10676 \\
1.39 & 2707 & 3291 & 5188 &   9247 \\
1.85 & 2270 & 2760 & 4351 &   7754 \\
2.59 & 1802 & 2191 & 3454 &   6156 \\
3.18 & 1548 & 1881 & 2966 &   5287 \\
  \hline
  \end{tabular}
 \medskip
 \\
 $^{a}$ the rest-frame wavelength (in $\rm \AA$) mapped
 by the central wavelength of each broad-band filter as a function of $z$
 \end{minipage}
\end{table}

\section{Results}

\subsection{Effects due to dust attenuation}

In this section we focus on the parameters describing the attenuation
by internal dust (see Sect. 2.2), an aspect neglected by studies
of high-$z$ galaxies that make use of the Calzetti law in a blind way.

Here we anticipate that the characteristics of dust attenuation
affect more the $\rm R_c$ magnitude than the K magnitude,
and turn out to be more sensitive parameters at higher redshifts,
since dust absorbs and scatters more at rest-frame UV--optical wavelengths
than at rest-frame near-IR ones, whatever its type (e.g. Whittet 2003).

\subsubsection{Effects due to structure of the dusty ISM, opacity, or inclination}

Figure 1 shows how the observed $\rm R_c$ and K magnitudes depend
on the structure (two-phase clumpy vs. homogeneous) assumed for the dusty ISM,
as a function of opacity, inclination, and redshift.
This is illustrated through models B1 and B2,
both with $\tau_{\star} \rm = 3~Gyr$, where the youngest stars
are the most embedded in a narrow dust lane.

A homogeneous dusty ISM is more ``opaque'' than a two-phase clumpy one
(e.g. Witt \& Gordon 1996, 2000).
However, the size of the relative absorption/scattering efficiency between
the two structures of a dusty ISM depends on opacity and inclination
of the disk, as quantified by Pierini et al. (2004b).
In fact, for the disk geometry, it is the total attenuation
along the observer's line of sight what counts.
In general, a face-on disk becomes more opaque at a given rest-frame wavelength
when the total amount of dust (i.e. the opacity) increases;
this is even more so if dust is distributed in a diffuse ISM.
Conversely, an edge-on disk becomes soon very opaque
when the opacity increases, so that differences in the structure
of its dusty ISM become unimportant.

For reference, Fig. 1 shows also the effect
of increasing the SFR e-folding time to 5 Gyr (i.e. models A1 vs. B1).
At lower redshifts (i.e. $z \la 1.4$) the observed optical/near-IR magnitudes
are more sensitive to the star-formation history than to the structure
of the dusty ISM.
The opposite is true for $z > 1.4$, especially if the disk model
has a large opacity (i.e. $2 \times \tau_{\rmn V} > 2$)
and is viewed at $i \la 70$\degr.
In general, as redshift decreases, star-formation effects are larger
on the $\rm R_c$ magnitude than on the K magnitude,
simply because the $\rm R_c$ band probes the light from the youngest stars.

\begin{figure}
 \vskip -1.0truecm
 \includegraphics[width=84mm]{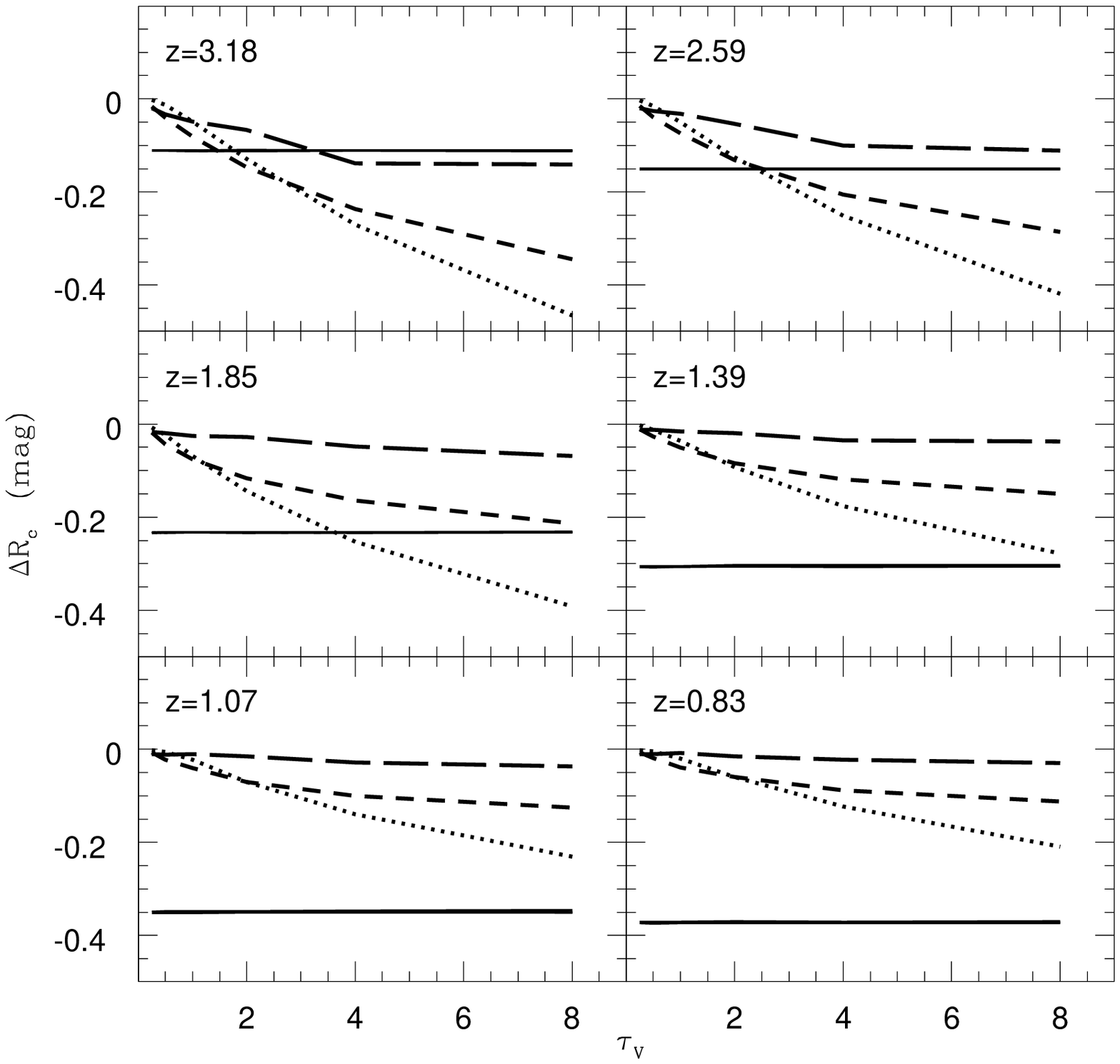}
 \vskip -1.5truecm
 \includegraphics[width=84mm]{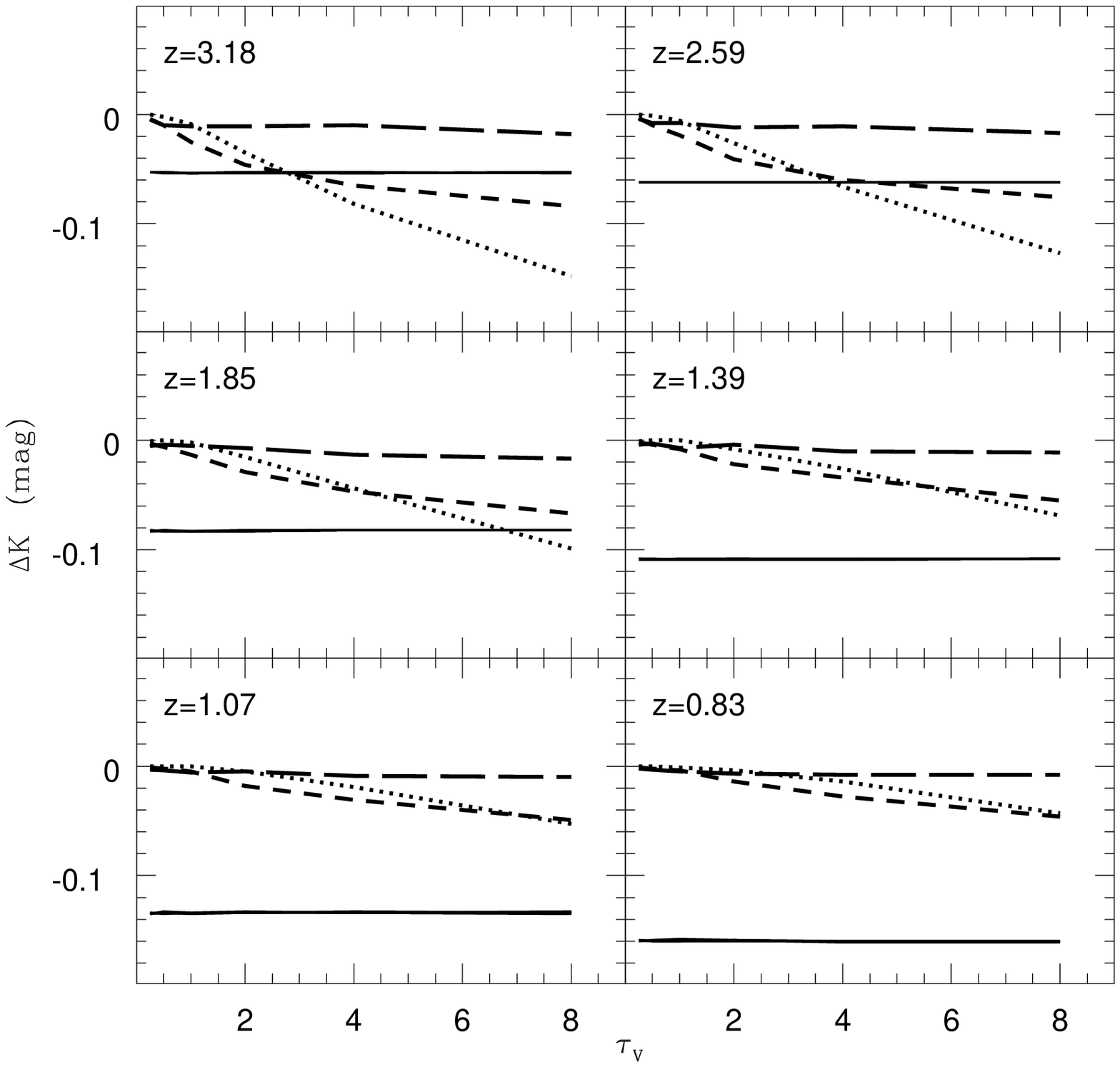}
 \caption{The difference ($\rm \Delta$) either in $\rm R_c$ magnitude (top)
  or in K magnitude (bottom) produced by a two-phase clumpy structure
  of the dusty ISM with respect to a homogeneous one (i.e. models B1 vs. B2)
  is plotted as a function of opacity ($2 \times \tau_{\rmn V}$),
  inclination ($i$), and redshift ($z$).
  Dotted, short-dashed, and long-dashed lines connect models
  with $i = 0$\degr, 70\degr, and 90\degr, respectively.
  Models B1 and B2 assume the youngest stars to be the most embedded
  in a narrow dust lane, and $\tau_{\star} \rm = 3~Gyr$.
  Solid lines show the magnitude differences when $\tau_{\star}$ increases
  to 5 Gyr (i.e. models A1 vs. B1).}
 \label{fig01}
\end{figure}

\subsubsection{Effects due to dust/stars configuration or dust type}

Figures 2 and 3 illustrate the sensitivity of the models
to dust/stars configuration and dust type,
for a fixed SFR e-folding time (i.e. 3 Gyr)
and for a dusty diffuse ISM (i.e. for models B3 through B8 in Tab. 1).
We recall that stars become more embedded in the dusty disk
when the dust/stars configuration changes from S01\_ME04 (or S01\_SE04)
to S01\_ME25 (or S01\_SE25).
In general, stellar light is attenuated and reddened more efficiently
when the dust-to-stars scale-height ratio increases from 0.4 to 2.5
(Ferrara et al. 1999).
Therefore, the kind of extinction law becomes more important
when the dust-to-stars scale-height ratio of the model increases,
opacity and inclination being fixed.

Differences in the observed $\rm R_c$ or K magnitude
due to a change in the dust/stars configuration increase (in absolute value)
when either the opacity or the inclination of the model increases,
whatever the dust type (see Fig. 2 for MW-type dust).
Interestingly, inclination effects are not negligible
and may not follow intuition.
For instance, an edge-on disk at high $z$ has a brighter $\rm R_c$ magnitude
when its stellar populations are fully embedded in the dusty ISM
(i.e. for the dust/stars configuration S01\_ME25 or S01\_SE25),
if $2 \times \tau_{\rmn V} < 2$.

The explanation of this apparent paradox is the following.
For an edge-on disk with $2 \times \tau_{\rmn V} < 2$, the dust column density
along the line of sight, averaged over the projection of the stellar disk
onto the sky plane, is lower the larger the dust-to-stars scale-height ratio.
In addition, scattering on dust grains is directed forward
at the rest-frame UV--blue wavelengths probed by the $\rm R_c$ band;
hence, more scattered UV--blue photons will leave the edge-on disk
at high $z$ when $2 \times \tau_{\rmn V} < 2$.
Conversely, for an edge-on disk with $2 \times \tau_{\rmn V} >> 2$,
the blocking action of the dusty ISM increases across a larger region
of the dusty disk overlapping with the stellar disk
the larger the dust-to-stars scale-height ratio.
A similar explanation applies to the behaviour of the observed K magnitude.
Dust grains not only absorb less but also scatter less and at a wide angle
at the rest-frame visual--near-IR wavelengths probed by the K band.
Hence, for an edge-on disk with $2 \times \tau_{\rmn V} < 2$,
more (rest-frame) visual--near-IR photons will be returned
to the observer's line of sight, traveling through the dusty disk
along different directions initially, the larger the dust-to-stars
scale-height ratio.

Figure 3 shows that a change in the extinction law has a relevant impact
on the $\rm R_c$ magnitude when the $\rm R_c$ band
maps rest-frame UV wavelengths.
For instance, for $z \sim 3.2$, SMC-type dust produces
a fainter $\rm R_c$ magnitude with respect to MW-type dust,
the relative dimming being of 0.6--0.8 mag at maximum,
as a function of inclination, for the models with dust-to-stars scale-height
ratio equal to 2.5 and $\tau_{\star} = 3$ Gyr reproduced there.
This differential effect is reduced for $z \sim 1.85$,
when the red-shifted $\rm 2175~\AA$-absorption feature of MW-type dust falls
inside the $\rm R_c$ band; it amounts up to 0.3 mag for lower redshifts
of the models.
SMC-type dust produces also a slightly fainter K magnitude,
by 0.25 mag at maximum, with respect to MW-type dust.
In conclusion, studies aimed at unraveling the extinction curves
of high-$z$ objects are urged (e.g. Hirashita et al. 2005).

As it can be inferred from Fig. 3, a change in the extinction law
produces a remarkable difference (up to 0.6 mag)
in the observed optical/near-IR colours of the models only when
the $\rm R_c$ and $\rm I_c$ broad-band filters map the rest-frame
spectral region around $\rm 2175~\AA$.
This happens for redshifts around $\sim 2$ and $\sim 3$ for the $\rm R_c$
and $\rm I_c$ filters, respectively.

\begin{figure}
 \vskip -1.0truecm
 \includegraphics[width=84mm]{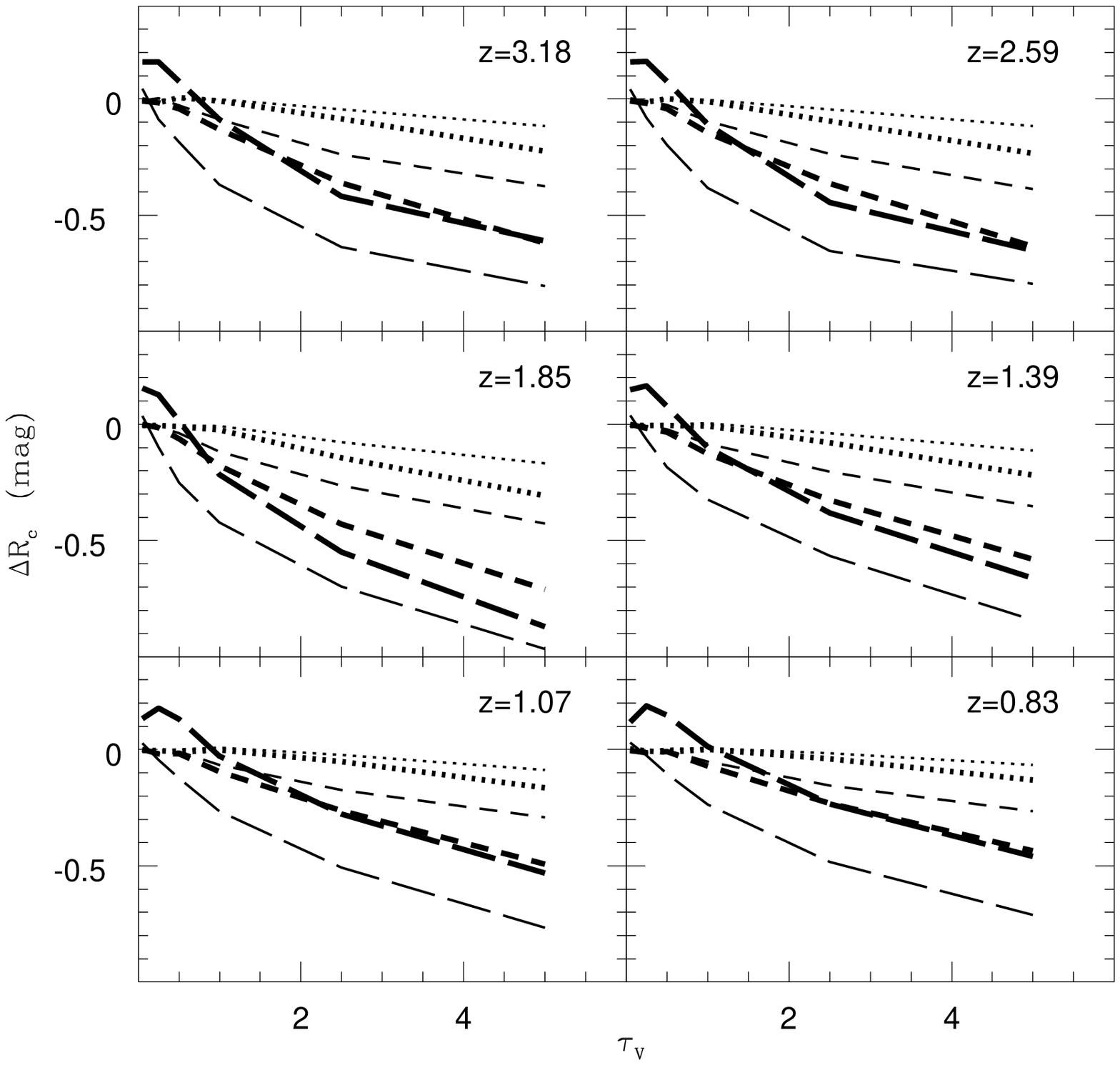}
 \vskip -1.5truecm
 \includegraphics[width=84mm]{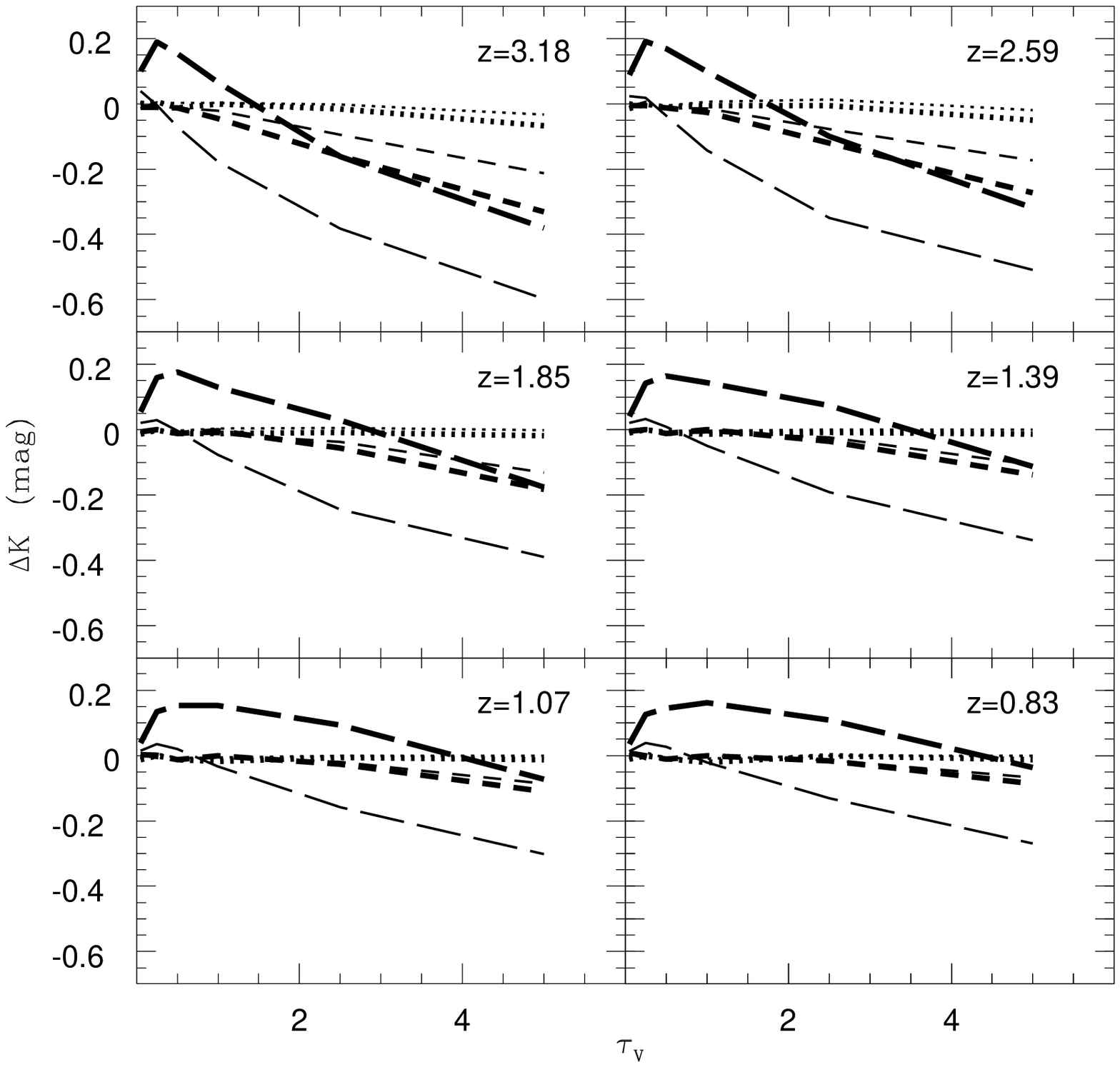}
 \caption{The difference either in $\rm R_c$ magnitude (top)
  or in K magnitude (bottom) between models B3 and B4 (thin lines)
  or models B3 and B5 (very thick lines) is plotted as a function of opacity,
  inclination, and redshift.
  Dotted, short-dashed, and long-dashed lines connect models
  with $i = 9.3$\degr, 70\degr, and 90\degr, respectively.  
  Models B3, B4, and B5 assume $\tau_{\star} \rm = 3~Gyr$,
  a diffuse ISM with MW-type dust,
  and a dust-to-stars scale-height ratio equal to 0.4, 1, and 2.5,
  respectively.}
 \label{fig02}
\end{figure}

\begin{figure}
 \vskip -1.0truecm
 \includegraphics[width=84mm]{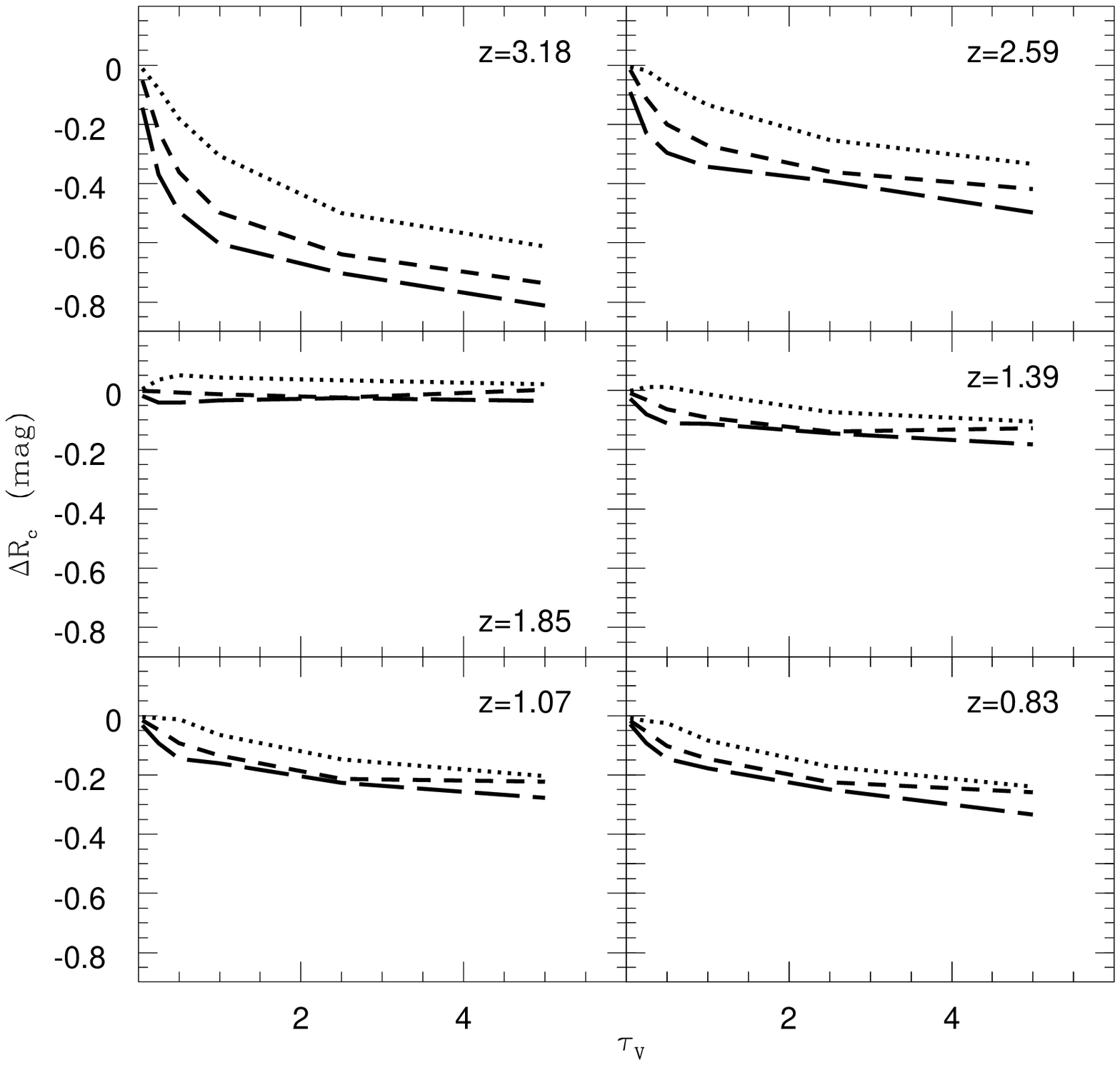}
 \vskip -1.5truecm
 \includegraphics[width=84mm]{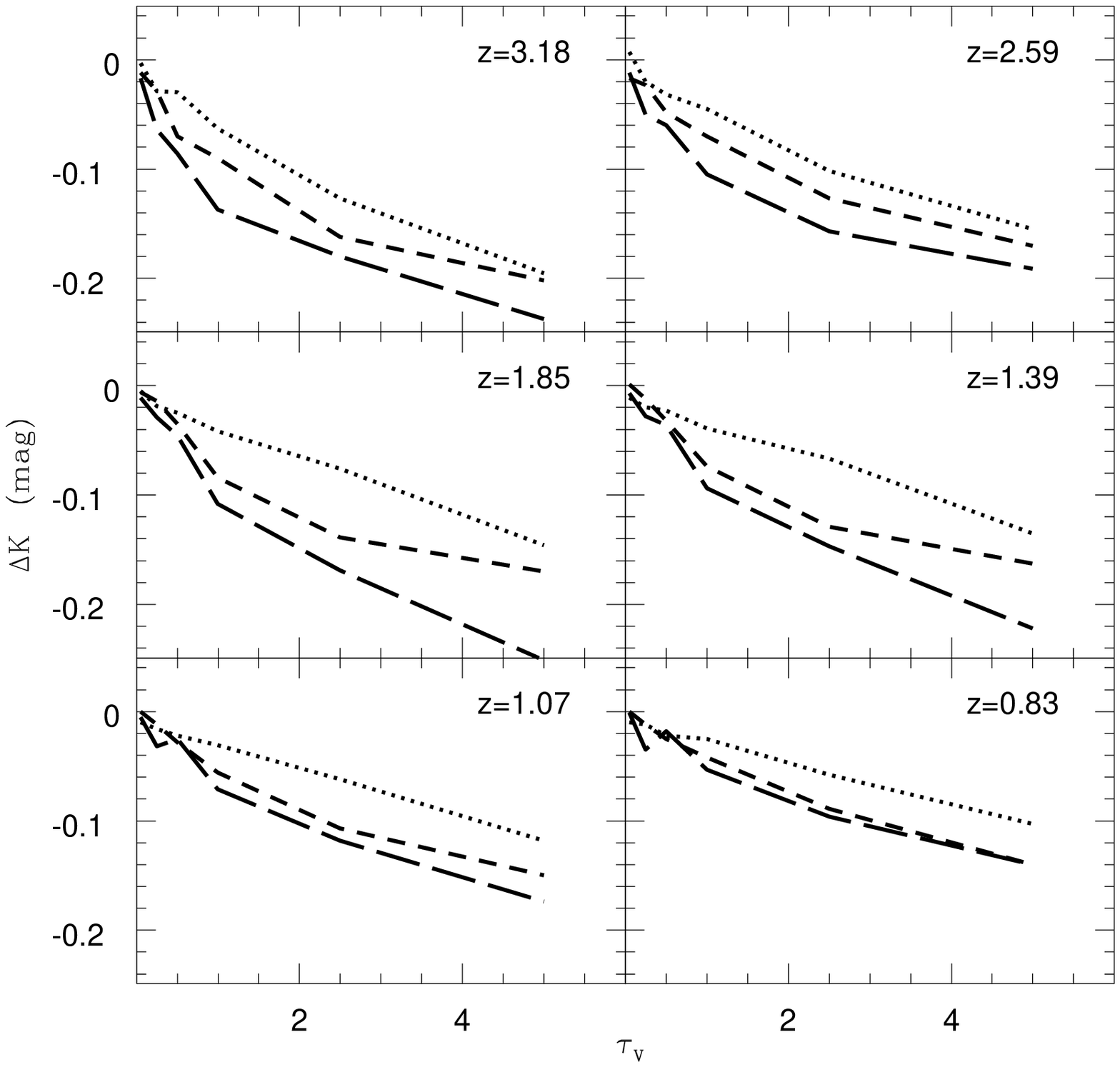}
 \caption{The difference either in $\rm R_c$ magnitude (top)
  or in K magnitude (bottom) produced by the MW extinction law
  with respect to the SMC one (i.e. models B5 vs. B8) is plotted
  as a function of opacity, inclination, and redshift.
  Dotted, short-dashed, and long-dashed lines connect models as in Fig. 2.
  Models B5 and B8 assume $\tau_{\star} \rm = 3~Gyr$, a dusty, diffuse ISM,
  and a dust-to-stars scale-height ratio equal to 2.5.}
 \label{fig03}
\end{figure}

\subsection{Models vs. observations}

\subsubsection{When high-$z$, dusty star-forming disks are ``red''}

\begin{figure}
 \vskip -1.0truecm
 \includegraphics[width=84mm]{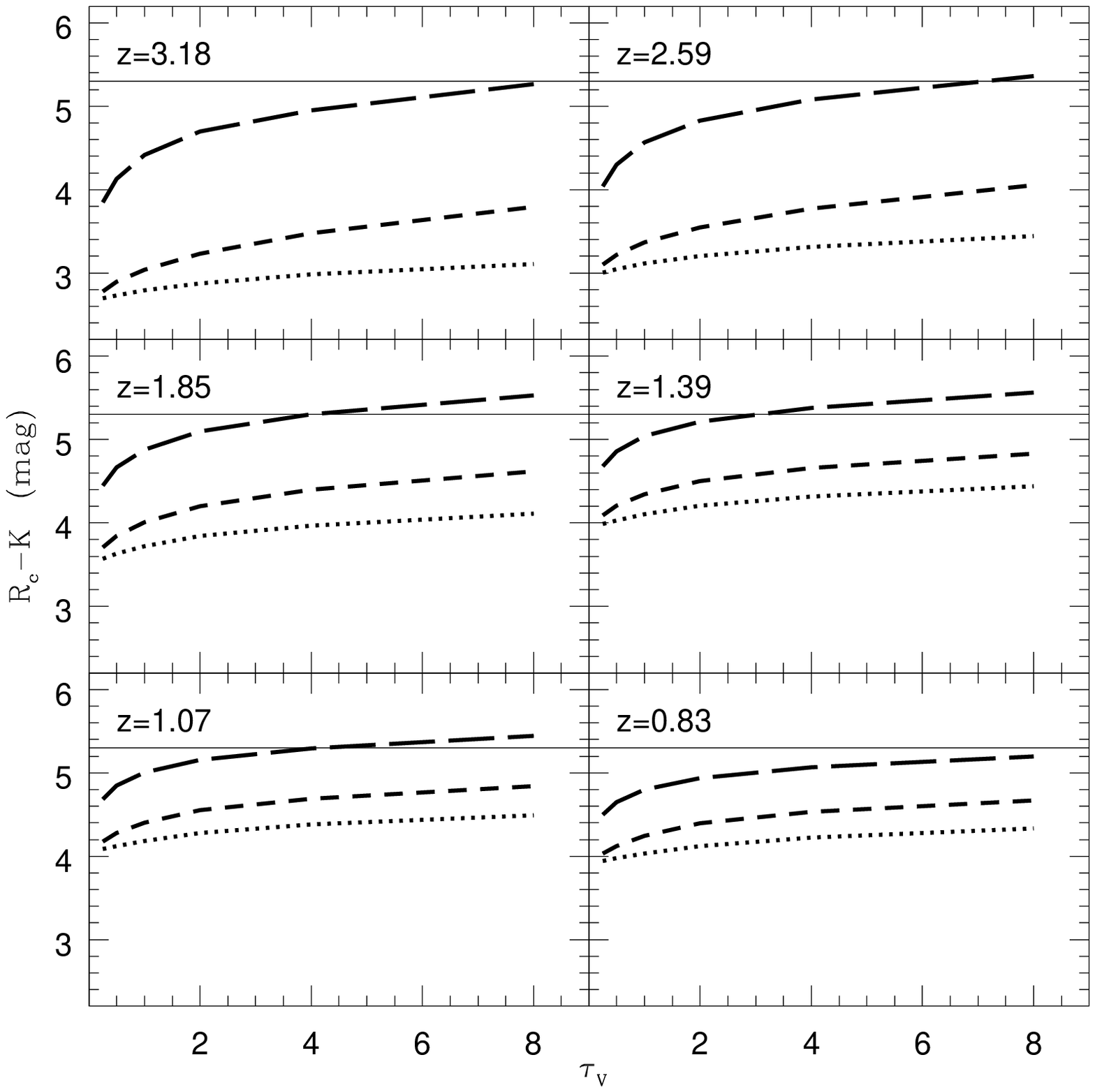}
 \vskip -1.5truecm
 \includegraphics[width=84mm]{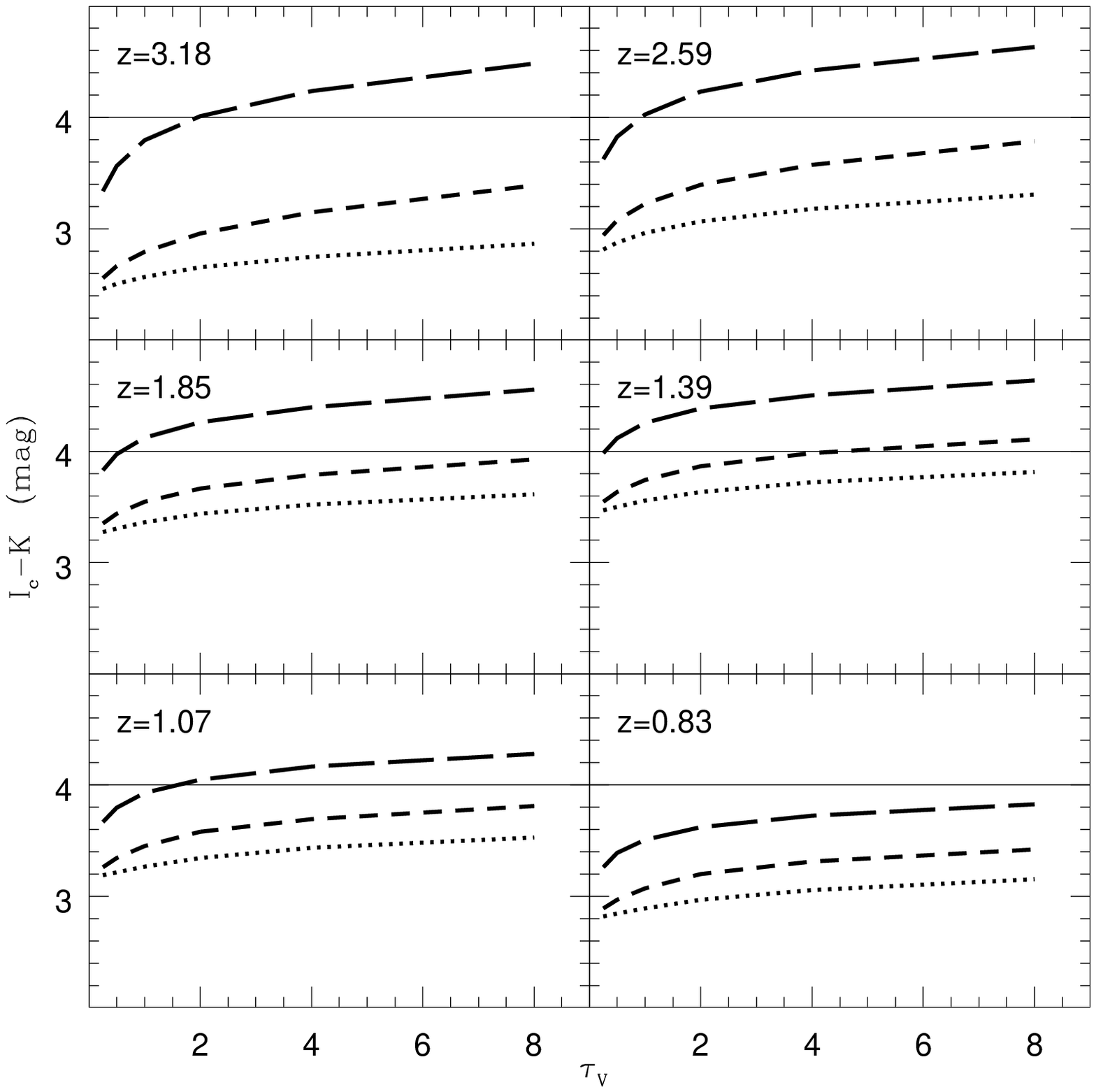}
 \vskip -1.5truecm
 \includegraphics[width=84mm]{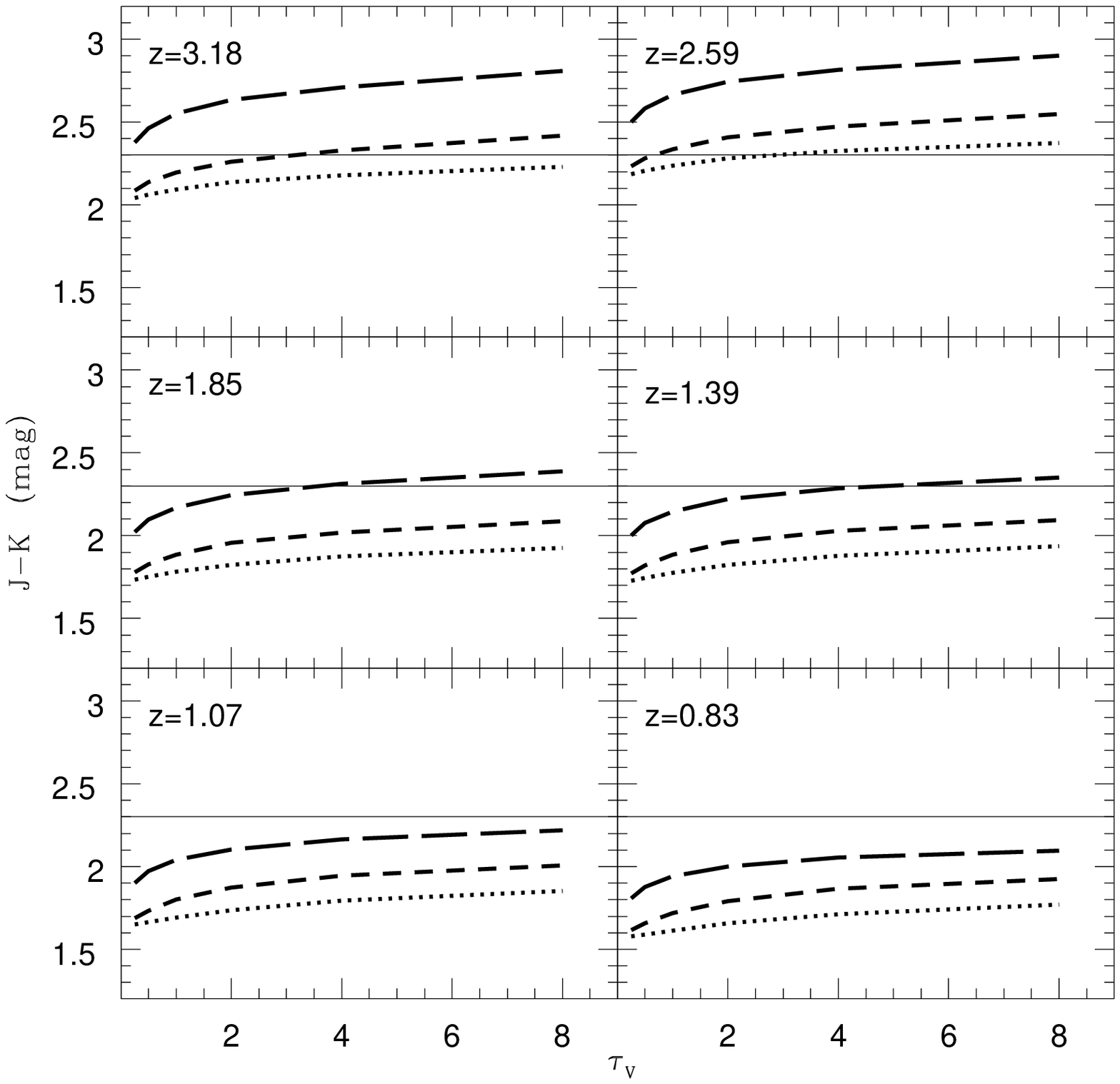}
 \caption{$\rm R_c - K$ (top), $\rm I_c - K$ (middle), and $\rm J - K$
  (bottom) are shown as a function of $\tau_{\rmn V}$, $i$, and $z$
  for models B1, where the youngest stars are the most embedded
  in a narrow lane with MW-type dust and a two-phase clumpy structure,
  and $\tau_{\star}$ is equal to 3 Gyr.
  Dotted, short-dashed, and long-dashed lines connect models
  with $i = 0$\degr, 70\degr, and 90\degr, respectively.
  Horizontal, solid lines show the different colour-selection criteria
  for high-$z$ red galaxies.}
 \label{fig04}
\end{figure}

\begin{figure}
 \vskip -1.0truecm
 \includegraphics[width=84mm]{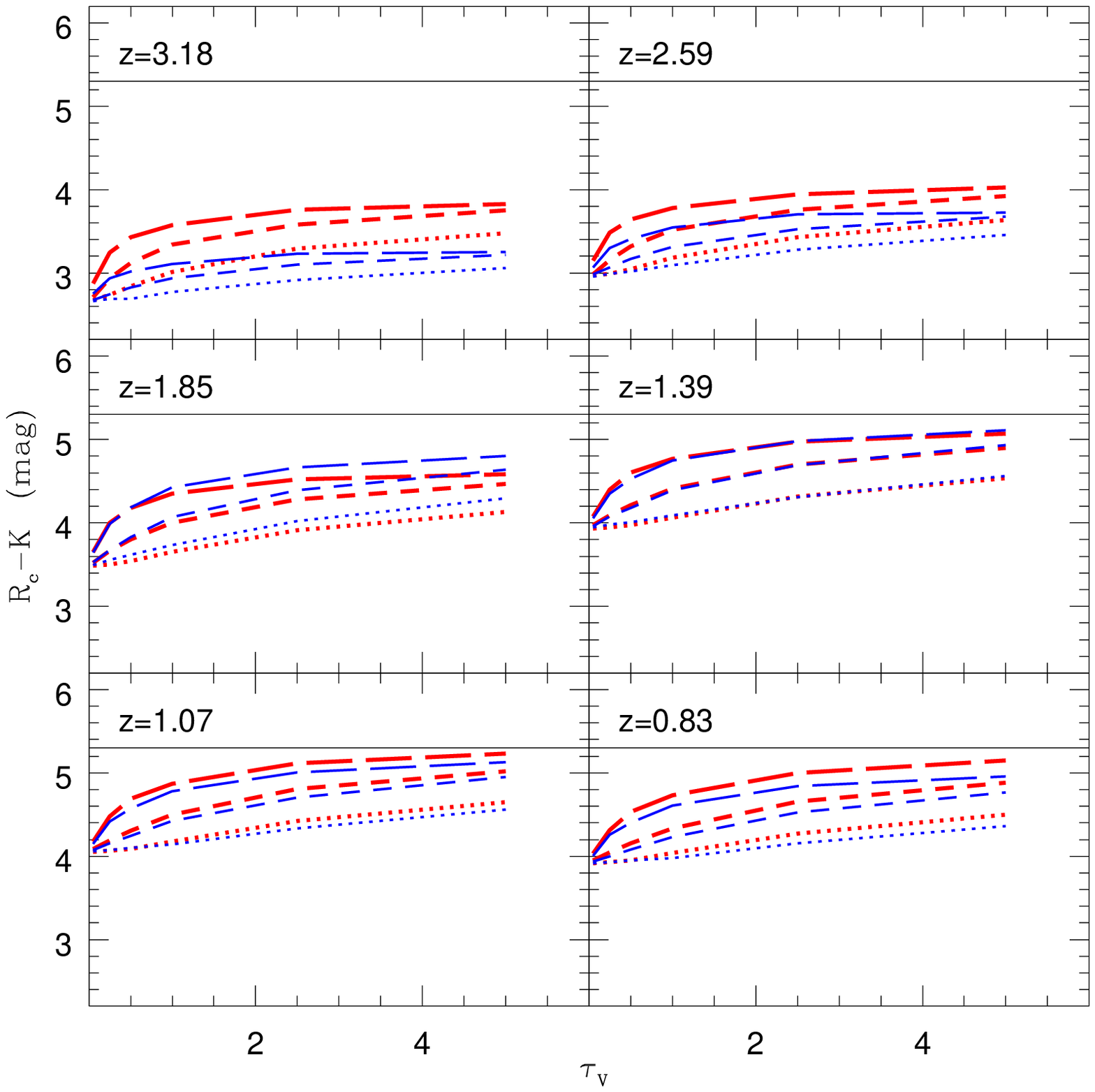}
 \vskip -1.5truecm
 \includegraphics[width=84mm]{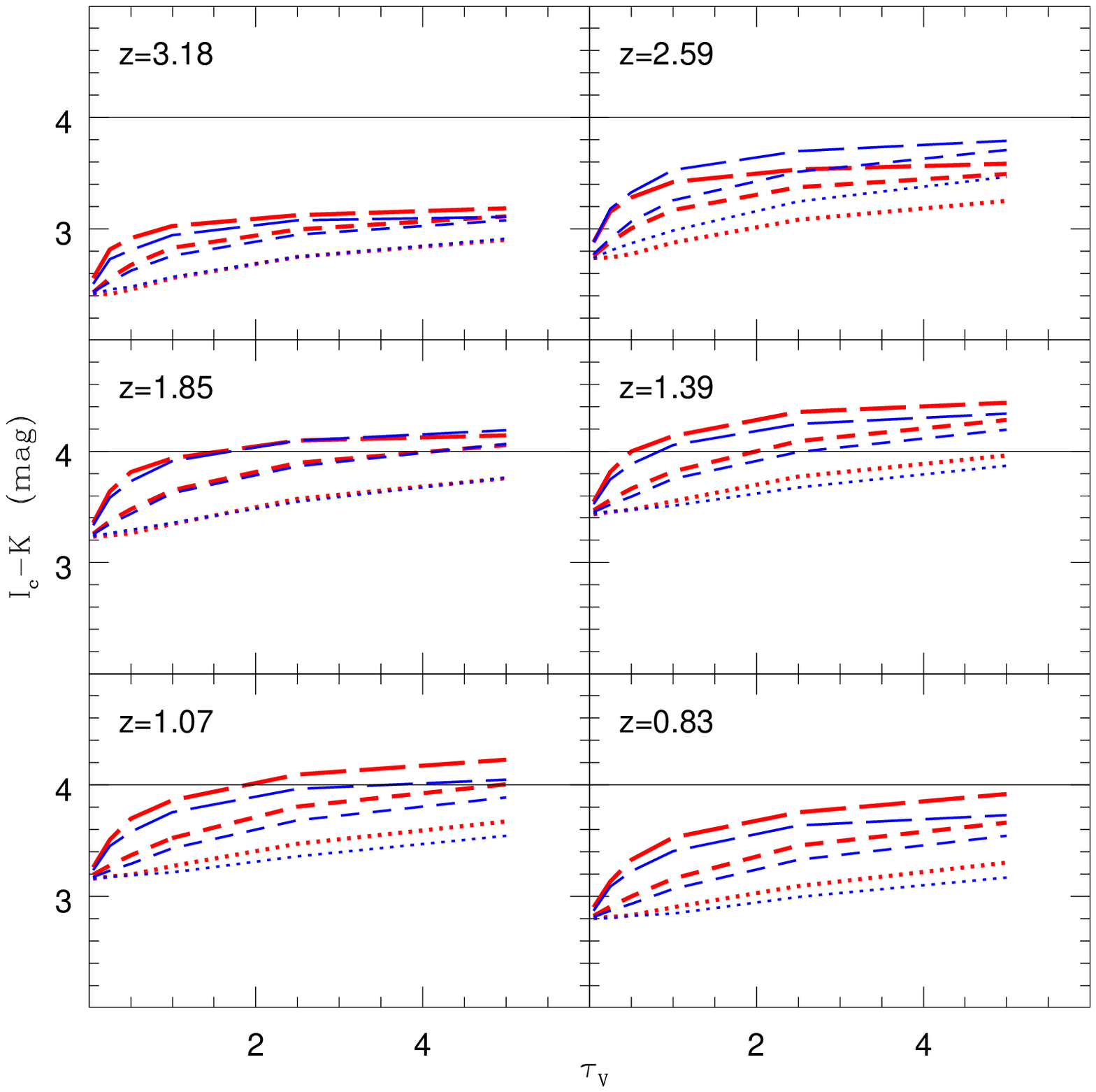}
 \vskip -1.5truecm
 \includegraphics[width=84mm]{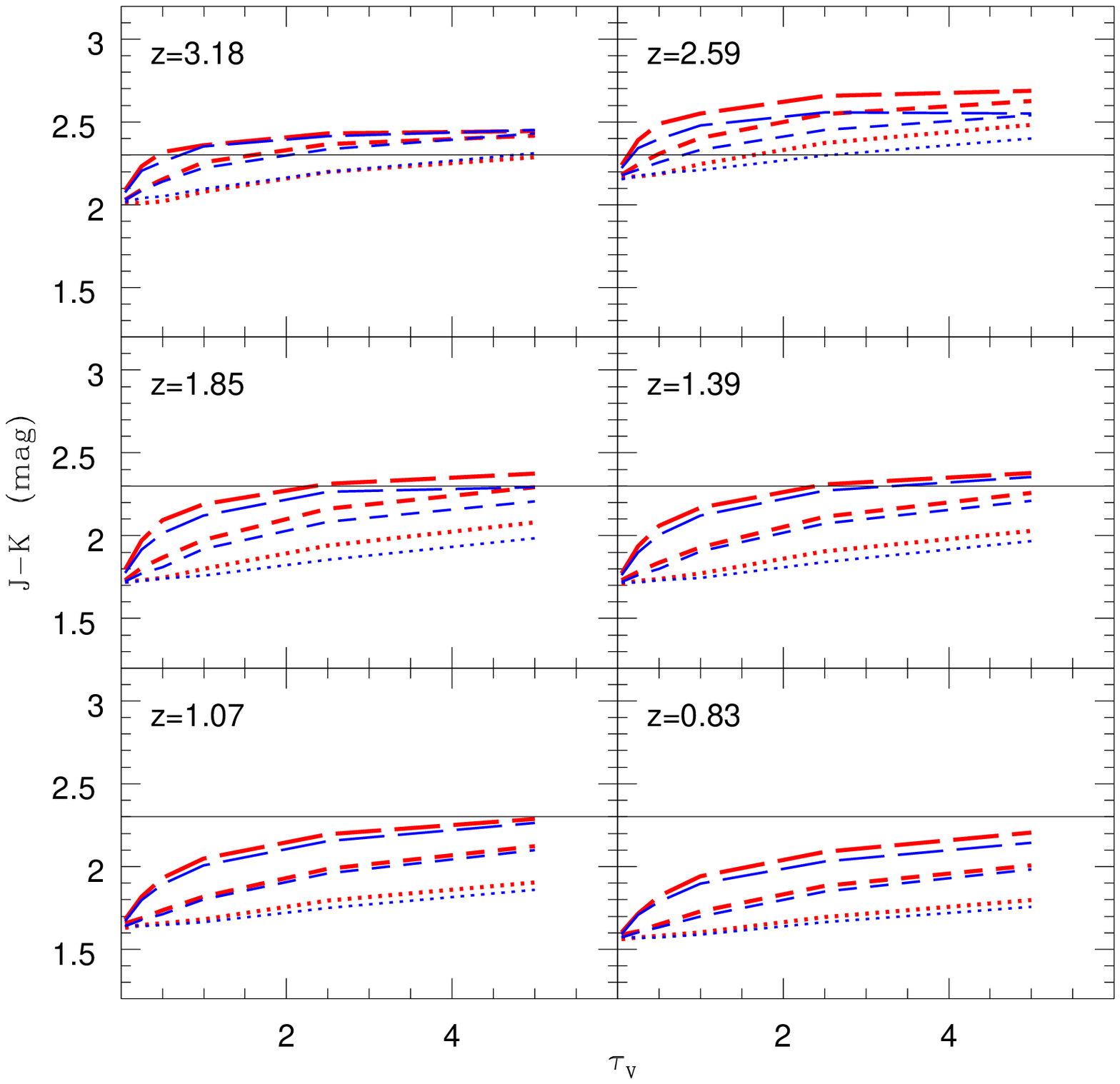}
 \caption{The same as in Fig. 4 but for models B5 (blue/thin lines) and B8
  (red/thick lines), where $\tau_{\star}$ is equal to 3 Gyr, the dusty ISM
  is diffuse, the dust-to-stars scale-height ratio is equal to 2.5,
  but dust is either of MW-type or of SMC-type, respectively.
  Dotted, short-dashed, and long-dashed lines connect models
  with $i = 9.3$\degr, 70\degr, and 90\degr, respectively.
  Horizontal, solid lines are the same as in Fig. 4.}
 \label{fig05}
\end{figure}

The results illustrated in Sect. 3.1.1 and 3.1.2 demonstrate that
the interpretation of the SEDs of high-$z$ galaxies
depends in a very sensitive way on the assumptions
made for describing dust attenuation.
Once a realistic description of dust attenuation in non-starburst
(disk) galaxies is adopted, do synthetic, high-$z$, dusty star-forming disks
exhibit the red colours identifying real DRdGs and ERdGs?
Hereafter we discuss which models are selected as DRdGs and/or ERdGs.
In particular, Fig. 4 and 5 illustrate results for models
with extreme properties as for the dust/stars configuration
and the extinction law (models B1, B5, and B8).

As a first main result, we find that the synthetic, dusty star-forming disks
that meet the colour selection criterion for distant red galaxies
(i.e. $\rm J - K > 2.3$) are mostly at $2 \la z \la 3.2$,
consistent with Franx et al. (2003);
no model at $z < 1$ exhibits $\rm J - K > 2.3$.
Models at $2 \la z \la 3.2$ can exhibit values of $\rm J - K$
as red as 3 mag at maximum, as it can be inferred from Fig. 4 and 5,
and from the considerations in Sect. 3.1.1 and 3.1.2.
Hence, they can account for a large fraction of the $\rm J_s - K_s$ colours
of the 34 real, distant red galaxies at (photometric) redshift
between 2 and 3.5 investigated by F\"orster Schreiber et al. (2004).
In fact, for this sample, the median, mean, and rms values of $\rm J_s - K_s$
are equal to 2.7, 2.8, and 0.43 mag, respectively.

However, our models can not reproduce $\rm J_s - K_s$ colours
in the range 3--3.9 mag, as observed for 12 of these 34 distant red galaxies.
Objects with $\rm J - K \ga 3$--4 were already observed
by Maihara et al. (2001).
These ``hyper extremely red objects'' have been proposed
to be primordial elliptical galaxies reddened by dust
and still in the starburst phase of their formation at $z \sim 3$
and, thus, to be the counterparts of the brightest sub-mm sources
(Totani et al. 2001).
However, a bulge$+$disk galaxy can offer a viable alternative
to the interpretation as starbursts of some of the previous 12
distant red galaxies at $2 \la z \la 3.5$ with $\rm 3 < J_s - K_s < 3.9$,
especially when considering that the properties of dust attenuation
are different between the bulge and disk components
(e.g. Pierini et al. 2004b).

In addition, we find that most of the $\rm J - K$-selected models
have $i \sim 90$\degr, despite the very different dust/stars configurations,
structures of the dusty ISM, and kinds of extinction law explored here.
The opacity of the $\rm J - K$-selected edge-on models at $2 \la z \la 3.2$
can be much less than 0.5.
On the other hand, a few $\rm J - K$-selected edge-on models have redshifts
as low as 1.4 if they are very opaque (e.g. $2 \times \tau_{\rmn V} \ga 8$
for models B1 in Fig. 4 or $2 \times \tau_{\rmn V} \ga 6$
for models B8 in Fig. 5).
However, synthetic $\rm J - K$-selected disks at $2 \la z \la 3.2$
can be viewed at inclinations as low as 0 degrees, if their opacity
is large enough (e.g. $2 \times \tau_{\rmn V} \ga 6$ for models B1 in Fig. 4
or $2 \times \tau_{\rmn V} > 4$ for models B8 in Fig. 5).

Unfortunately, there is almost no morphological information
for the F\"orster Schreiber et al. (2004) sample at the moment of writing,
except for the DRdG discovered by Labb\'e et al. (2003),
exhibiting $\rm J_s - K_s = 2.6$ and $\rm F814W - K_s = 3.9$.
In order to reproduce the optical/near-IR colours
of this large disk-like galaxy at $z = 2.94$ (photometric),
seemingly viewed at $i < 70$\degr, with e.g. a dust/stars configuration
like that of Pierini et al. (2004a), one has to invoke
$\tau_{\star} \rm < 3~Gyr$.
Hence, this object might be either experiencing or on the verge of
a central gas$+$stars instability (see Immeli et al. 2004).
Similar conclusions may apply to the disk-like galaxy at $z \sim 2.5$
(photometric) with $\rm J - K^{\prime} \sim 3.4$ and $i \sim 70$\degr
discovered by Stockton et al. (2004).

Interestingly, Toft et al. (2005) conclude that
the rest-frame UV and optical surface brightness distributions
of their 5 $\rm J - K_s$-selected galaxies at $z \ga 2$ (photometric)
are better represented by exponential disks than by $r^{1/4}$ laws.
Two of these disk galaxies (at $z = 2.2$ and 1.9) certainly host
active galactic nuclei (AGN, see also Rubin et al. 2004)
and, thus, have bulges, if the link beetwen super massive black holes
and bulges found in the local Universe holds at high redshifts
(Page et al. 2004).
Their $\rm J - K_s$ colours (2.42 and 2.66 mag) are not particularly red
however: the high-inclination disk-like galaxy at $z = 1.8$
of Toft et al. (2005) exhibits $\rm J - K_s = 2.45$.

In conclusion, our $\rm J - K$-selected models of dusty, star-forming,
bulge-less disk galaxies do not necessarily suffer
from large rest-frame V-band attenuations.
This conclusion holds even though a decrease of $\tau_{\rmn V}$ by a factor
of about 2 can be produced by an equivalent decrease of $\tau_{\star}$,
all the other model parameters being fixed (not shown).
It also holds whatever the dust/stars configuration,
the structure of the dusty ISM, or the dust type,
for fixed star-formation history\footnote{The simultaneous presence
of SMC-type dust and of a geometrically-thin dust lane where the youngest stars
are the most embedded maximizes the fraction of successful models.}.

This result is at variance with that of F\"orster Schreiber et al. (2004),
in case most of their observed distant red galaxies are indeed
bulge-less disk galaxies at $2 \la z \la 3.5$.
For instance, 2 Gyr-old (at $z = 2.59$) models B1
with $\tau_{\star} \rm = 3~Gyr$ exhibit $\rm 2.3 < J - K < 2.5$
already for $i \sim 70$\degr, if $2 \times \tau_{\rmn V} > 1$,
i.e. for a rest-frame V-band attenuation along the line of sight
larger than $\rm \sim 0.2~mag$, and $\rm 2.3 < J - K < 2.4$
for $i \sim 0$\degr, if $2 \times \tau_{\rmn V} \ga 6$,
i.e. for a rest-frame V-band attenuation along the line of sight
larger than $\rm \sim 0.3~mag$ (see Fig. 4).
Conversely, the model-fitting technique adopted by F\"orster Schreiber et al.
gives median ages equal to 1 or 2 Gyr and, respectively, 
median values of $A_{\rmn V}$ equal to 0.9 or 2.5 mag
when the SFR is assumed either to decline exponentially with time as in Eq. 1
but with $\tau_{\star} \rm = 0.3~Gyr$ or to be constant.

The discrepant conclusions can be explained as follows.
Firstly, a realistic description of dust attenuation for the disk geometry
(but not only, see e.g. Witt \& Gordon 2000), as from radiative transfer
calculations, shows that the ratio of UV-to-visual attenuation
changes significantly and non monotonically
as a function of the V-band attenuation (e.g. Ferrara et al. 1999;
Tuffs et al. 2004; Pierini et al. 2004b).
Conversely, this ratio is ``frozen'' in the Calzetti law adopted
by F\"orster Schreiber et al. (2004),
as shown by e.g Pierini et al. (2004a, their fig. 2).

Secondly, the TP-AGB phase included in the stellar population models
impacts on their rest-frame visual--near-IR SEDs (see Maraston 2005)
in a way that produces intrinsically redder, observed near-IR colours
for $2 \la z < 3.2$.
F\"orster Schreiber et al. (2004) resort to the Bruzual \& Charlot (2003)
stellar population evolutionary synthesis models that do not include
the TP-AGB phase (see Maraston 2005).
We have evaluated the reddening produced by the TP-AGB phase
(up to $\sim 0.2$ mag in the observed $\rm J - K$ and $\rm I_c - K$,
and up to $\sim 0.3$ mag in the observed $\rm R_c - K$)
by computing equivalent models in which the TP-AGB phase
was neglected on purpose.
The effect of the TP-AGB phase depends obviously on the redshift,
because it is necessary that the observed frame probes the near-IR rest-frame
where the TP-AGB stars emit most of their energy (see Sect. 2.1).
Moreover, the ratio between the numbers of old and intermediate-age stars
increases together with the age (i.e. with decreasing redshift).
As a consequence, the impact of the emission from TP-AGB stars
on the composite stellar population models considered here is maximum
for $1 \la z \la 2$, when the age of the models approaches
the SFR e-folding time.

In conclusion, the effect of TP-AGB stars is to allow more models
of dusty, star-forming disks at $0.8 \la z \la 3.2$ to be selected
as either DRdGs or ERdGs, and to reduce the needed amount of dust
by up to a factor of two (see Fig. 6).

As for ERdGs, the $\rm I_c - K$-selected models at $1 \la z \la 2$
need to have an inclination of 90\degr, but in case the dusty ISM is diffuse
and the dust-to-stars scale-height ratio is equal to 2.5,
$i \sim 70$\degr is sufficient (see Fig. 5).
The synthetic, $\rm I_c - K$-selected edge-on disks exhibit a large range
in opacity (i.e. $2 \times \tau_{\rmn V} \ga 0.5$),
according to the dust/stars configuration, the structure of the dusty ISM,
the dust type, and the star-formation history.
The value of the rest-frame V-band attenuation
of the $\rm I_c - K$-selected edge-on models is very sensitive
to all these parameters.
For instance, edge-on models A1 at $z = 1.39$ exhibit $\rm 4 < I_c - K \la 4.5$
for $2 \times \tau_{\rmn V} > 1$, i.e. for values of the rest-frame V-band
attenuation along the line of sight larger than $\rm \sim 0.8~mag$.
Reducing $\tau_{\star}$ from 5 to 3 Gyr makes the edge-on models B1
at $z = 1.39$ span a similar colour range ($\rm 4 < I_c - K \la 4.6$)
but already for $2 \times \tau_{\rmn V} > 0.5$, i.e. for values
of the rest-frame V-band attenuation along the line of sight
larger than $\rm 0.5~mag$.
Conversely, edge-on models B3 and B6, with a dust-to-stars scale-height ratio
equal to 0.4, never exhibit $\rm I_c - K > 4$.
We conclude that real $\rm I_c - K$-selected, edge-on disk-like galaxies
at $1 \la z \la 2$ are not necessarily more opaque, and, thus, dustier
than nearby disk galaxies, for which $0.5 \la 2 \times \tau_{\rmn V} \la 2$
(Kuchinski et al. 1998).

Far-IR photometry of high-$z$ red disk-like galaxies with {\it Spitzer}
and {\it Herschel} will offer a decisive test for this conclusion.
This holds for DRdGs at $2 \la z \la 3.2$ as well.

\begin{figure}
 \vskip -1.0truecm
 \includegraphics[width=84mm]{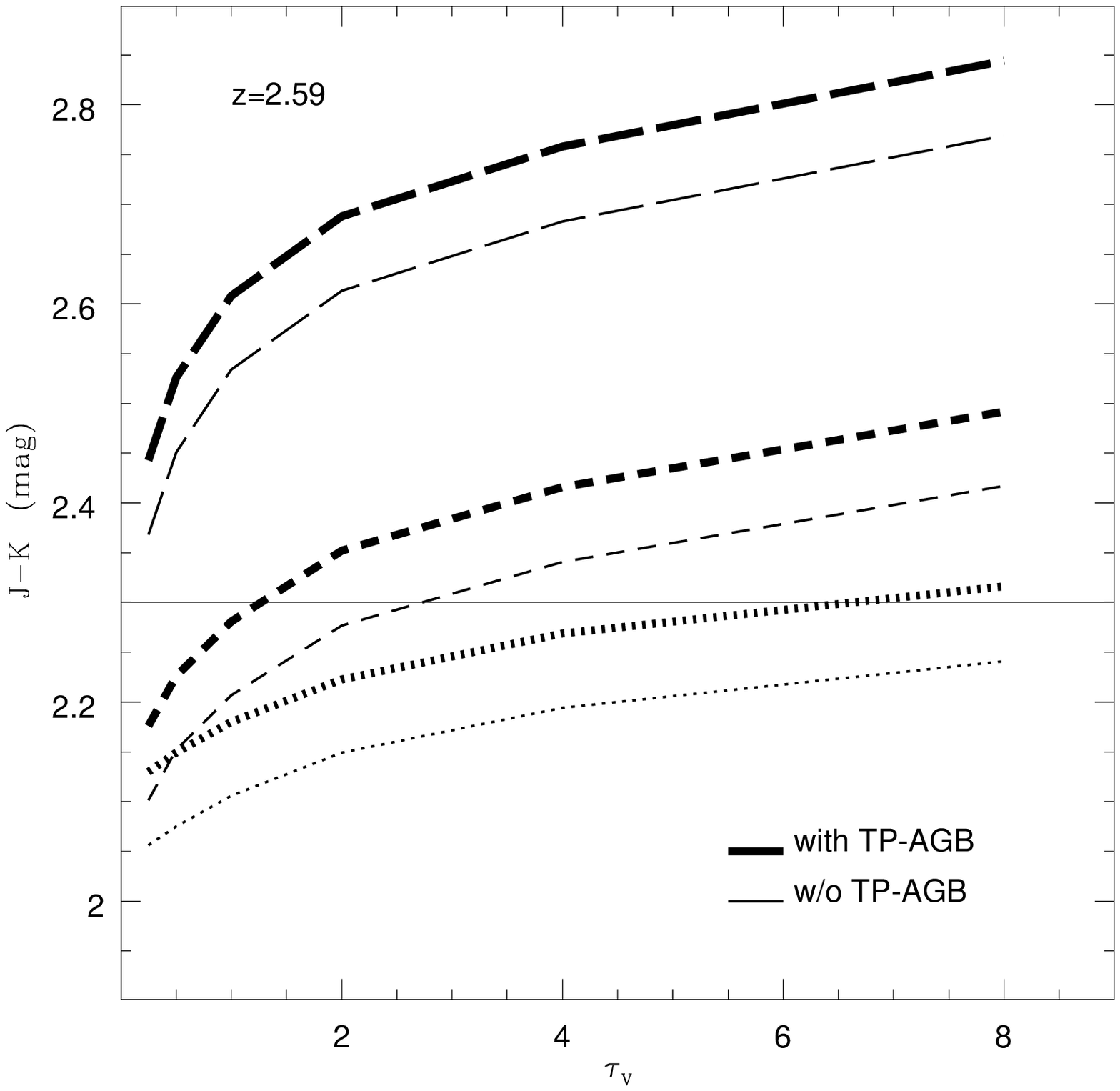}
 \vskip -1.5truecm
 \includegraphics[width=84mm]{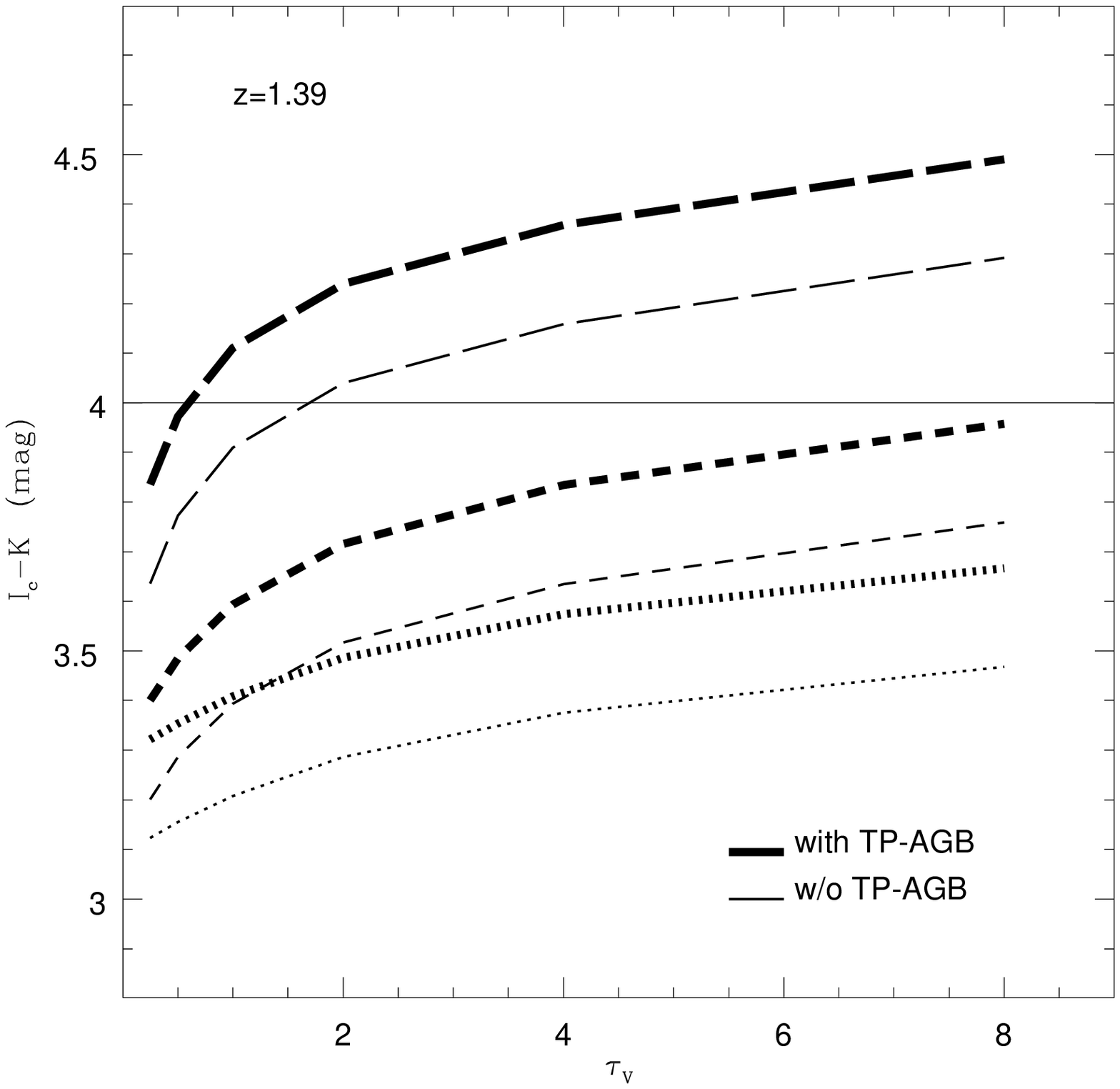}
 \caption{$\rm J - K$ (top) and $\rm I_c - K$ (bottom) are plotted
  as a function of $\tau_{\rmn V}$ and $i$ for models A1 at, respectively,
  $z = 2.59$ and $z = 1.39$ (thick lines) and for equivalent models
  in which the TP-AGB phase was neglected on purpose (thin lines).
  Dotted, short-dashed, and long-dashed lines connect models
  with $i = 0$\degr, 70\degr, and 90\degr, respectively.
  Horizontal, solid lines show the appropriate colour-selection criterion.}
 \label{fig06}
\end{figure}

The previous result is new and encouraging for the following reasons.
Firstly, the unspecified model SEDs for Sb-type galaxies adopted
by V\"ais\"anen \& Johansson (2004) from the GRASIL model library
(Silva et al. 1998) failed to reproduce the optical/near-IR colours
identifying extremely red galaxies, that do contain late-type galaxies
(Yan \& Thompson 2003; Cimatti et al. 2003; Gilbank et al. 2003).
This is most probably due to the use of angle averaged, standard GRASIL models
for Sb galaxies, where dust and stars have the same spatial distribution
(see Silva et al. 1998).

Secondly, and most importantly, HST imaging in the $\rm F814W$ broad-band
filter of $\rm F814W - K_s$-selected galaxies with $\rm K_s < 19.5$ shows that
one third of this sample is made of disky systems (i.e. disk-dominated
or bulge-less) viewed almost edge on (Yan \& Thompson 2003).
Moreover, a few edge-on ERdGs exhibit narrow dust lanes beyond any doubt,
thanks to the superb spatial resolution of HST.
Now, Yan \& Thompson conclude that most of their edge-on disky systems
lie possibly at $z < 1$, otherwise these systems would be too large.
However, bright, $\rm F814W - K_s$-selected disk-like galaxies do exist
at $z \sim 1$ (Yan et al. 2004) and our models can explain this finding.

At this point we note that 3 out of the 7 $\rm F814W - K_s$-selected,
disk-dominated galaxies with spectroscopic redshifts around 1 and classified
as absorption-line objects by Yan et al. (2004) are viewed close to edge on,
and 6 out of the 13 objects without measured redshifts are also disk-dominated
galaxies seen edge on.
Yan et al. suggest that the absence of detected spectral features
(in the observed range 5500--$\rm 9000~\AA$) may be due to their weakness,
owing to dust attenuation (emission lines) and/or to the intrinsic mixture
of stellar populations (emission and absorption lines).
Our result confirms that weak spectral features may be due
to an increased contribution of intermediate-age stellar populations
to the total bolometric luminosity of the ensamble of stars
present in the previous $\rm F814W - K_s$-selected disk-dominated galaxies,
without invoking extremely large amounts of dust.

Finally, only a handful models can meet the $\rm R_c - K > 5.3$
colour-selection criterion for extremely red galaxies at $1 \la z \la 2$.
They all have a geometrically-thin dust lane where the youngest stars
are the most embedded in, a very large opacity
(i.e. $2 \times \tau_{\rmn V} \ga 8$), and an inclination of 90\degr.
No unambigous discovery of $\rm R_c - K$-selected, bulge-less
disk galaxies at redshifts between 1 and 2 is reported in the literature
(see e.g. Sawicki et al. 2005).
The absence of continuously star-forming, bulge-less disk galaxies
at $1 \la z \la 2$ in $\rm R_c - K$-selected samples is due to
a selection effect, the $\rm R_c - K > 5.3$ threshold being too red
for such objects (as explained in Sect. 3.2.8).

\subsubsection{On the epoch of initial star formation}

Figure 7 shows the distribution in the $\rm R_c - K$ vs. $\rm J - K$
and $\rm I_c - K$ vs. $\rm J - K$ colour--colour plots of models A1,
with ages between 1.5 and 6 Gyr for $z$ between 3.2 and 0.8, respectively,
and for 0.6 Gyr-old models of dusty star-forming disks at $0.8 \la z \la 3.2$
(models C1).
For these two sets of models with $\tau_{\star} \rm = 5~Gyr$,
the dust/stars configuration, the local dust distribution,
and the dust type are the same.

In order to reproduce red near-IR colours (i.e. $\rm J - K > 2.3$)
or extremely red optical/near-IR colours (i.e. $\rm I_c - K > 4$),
the time that has elapsed since star formation started is at least $\sim 1$ Gyr
at the redshift of the colour-selected model.
Otherwise, either dust attenuation for the high-$z$ objects is not described
as for disk galaxies in the local Universe, or bulge-less disk galaxies
with star-formation time-scales of the order of $\sim 1$ Gyr do exist
at high $z$.

Comfortably, evolved stellar populations seem to be present
in bright, $\rm F814W - K_s$-selected, edge-on disk-dominated galaxies
classified as absorption-line systems (Yan et al. 2004).

On the other hand, also F\"orster Schreiber et al. (2004) conclude
that the median time that has elapsed since star formation started
ranges between 1 and 2 Gyr for their sample of $\rm J - K$-selected galaxies
at $2 \la z \la 3.5$ (photometric).
Therefore, our conclusion on the formation epoch of DRdGs at these redshifts
is consistent with that of F\"orster Schreiber et al., though we fear that
the assumptions behind their models may lead to an artificial increase
of the ages and/or the values of the rest-frame V-band attenuation
given by their best-fit solutions in case of disks (see Sect. 3.2.1).
The same applies to the model-fitting results of Toft et al. (2005).

\begin{figure}
 \vskip -1.0truecm
 \includegraphics[width=84mm]{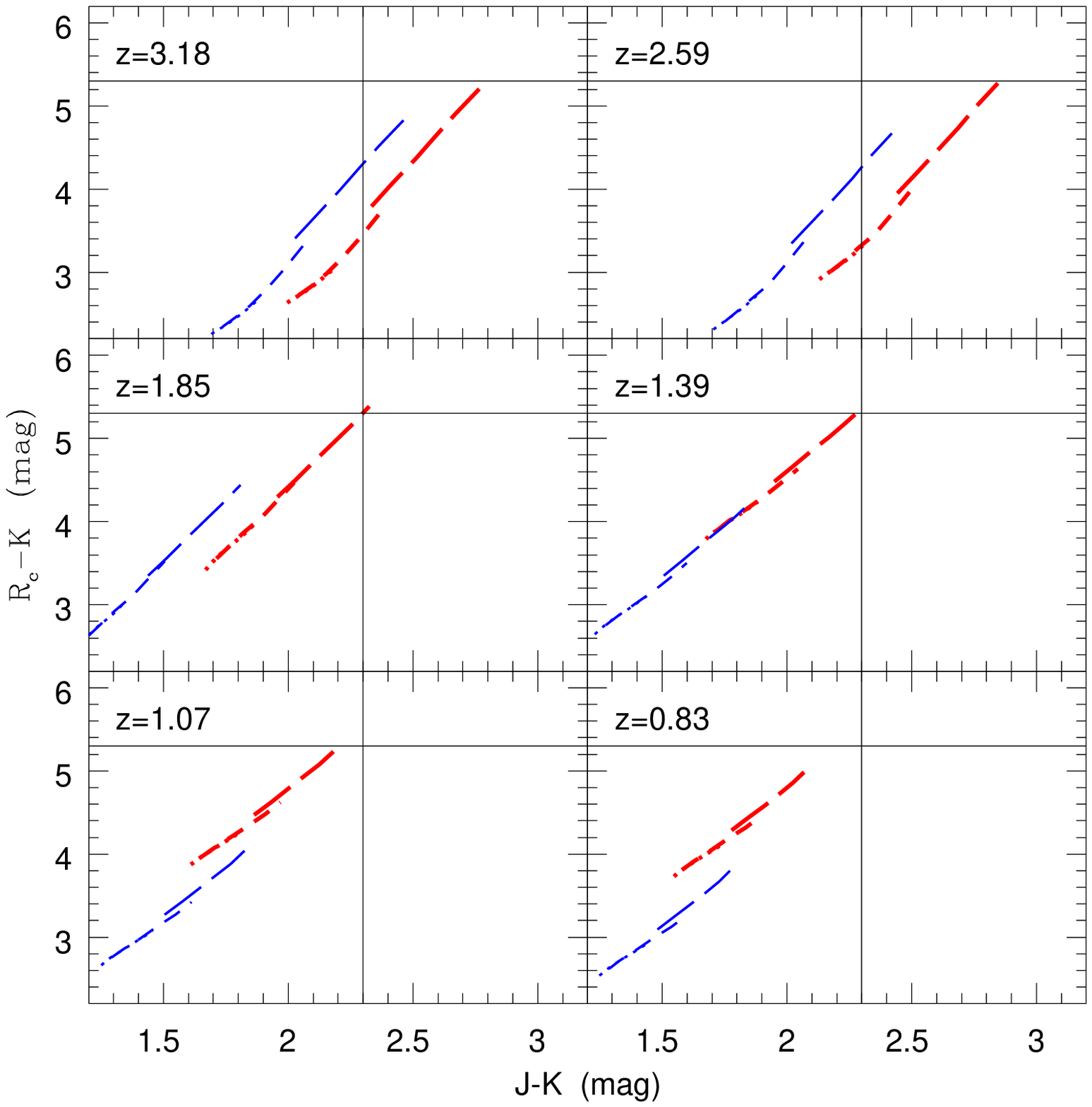}
 \vskip -1.5truecm
 \includegraphics[width=84mm]{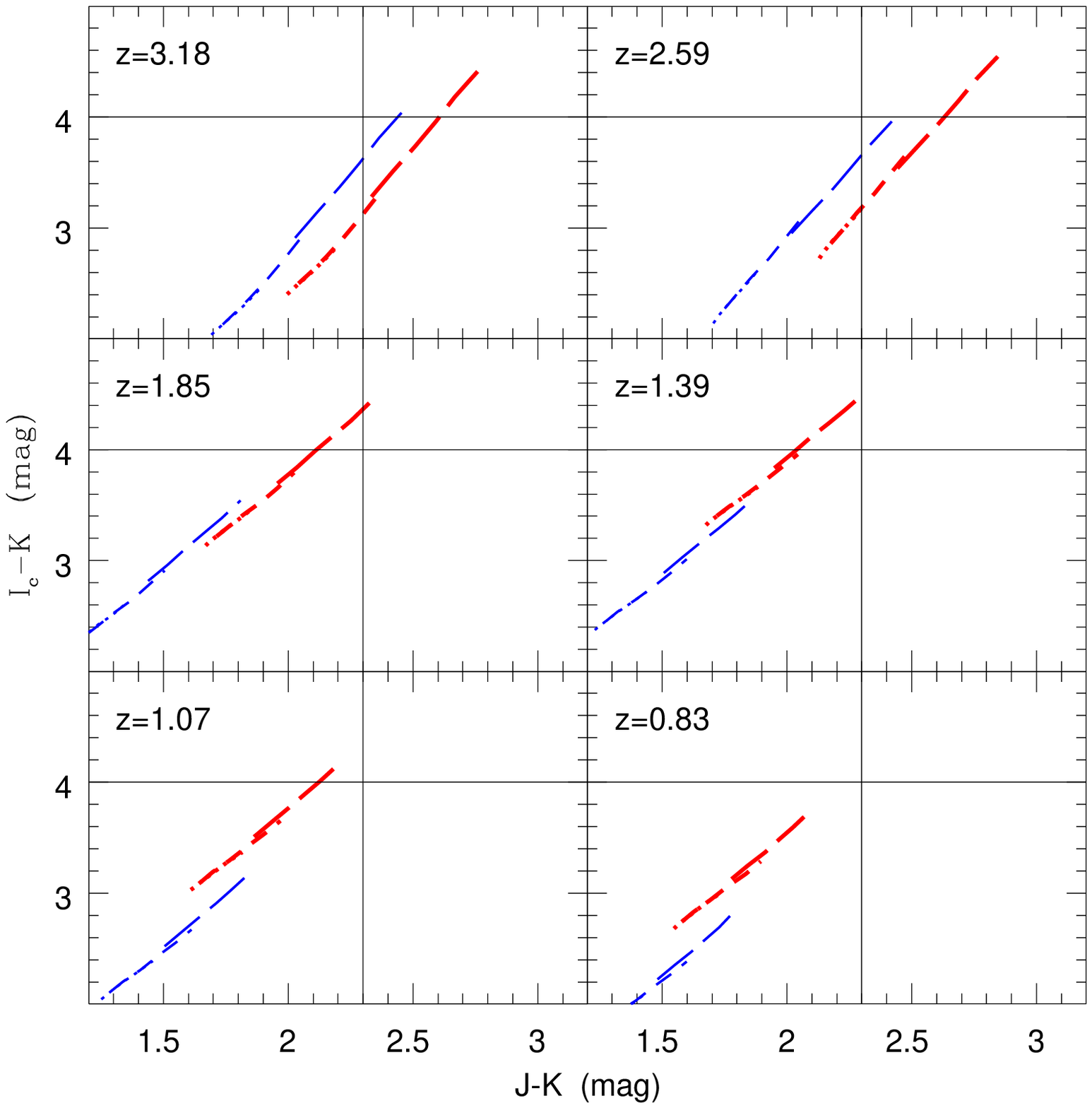}
 \caption{$\rm R_c - K$ vs. $\rm J - K$ (top) and $\rm I_c - K$ vs. $\rm J - K$
  (bottom) are shown as a function of opacity, inclination, and redshift
  for models C1 (blue/thin lines) and A1 (red/thick lines),
  where the youngest stars are the most embedded in a narrow lane
  with MW-type dust and a two-phase clumpy structure,
  and $\tau_{\star} \rm = 5~Gyr$.
  However, models C1 are 0.6 Gyr old whatever $z$,
  while models A1 age with decreasing redshift, having $z_{\rmn f} = 10$.
  Dotted, short-dashed, and long-dashed lines connect models
  with $i = 0$\degr, 70\degr, and 90\degr, respectively; for each line,
  the opacity of the model increases from bottom left to top right.
  In each panel, horizontal and vertical, black solid lines show
  the colour selection criteria for extremely red galaxies
  and distant red galaxies, respectively.}
 \label{fig7}
\end{figure}

\subsubsection{The co-moving space density of high-$z$ red disk-like galaxies}

\begin{figure}
 \vskip -1.0truecm
 \includegraphics[width=84mm]{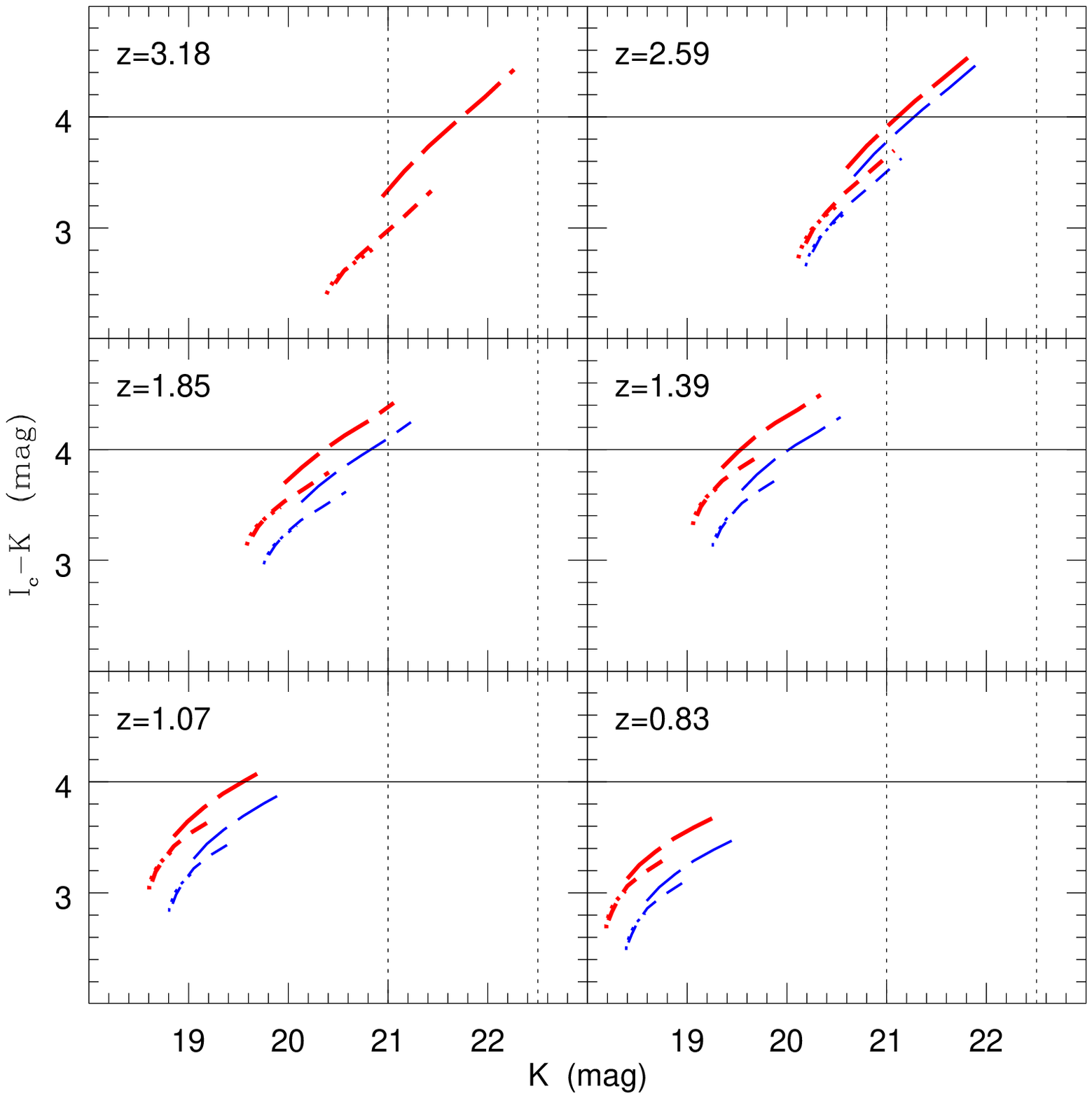}
 \vskip -1.5truecm
 \includegraphics[width=84mm]{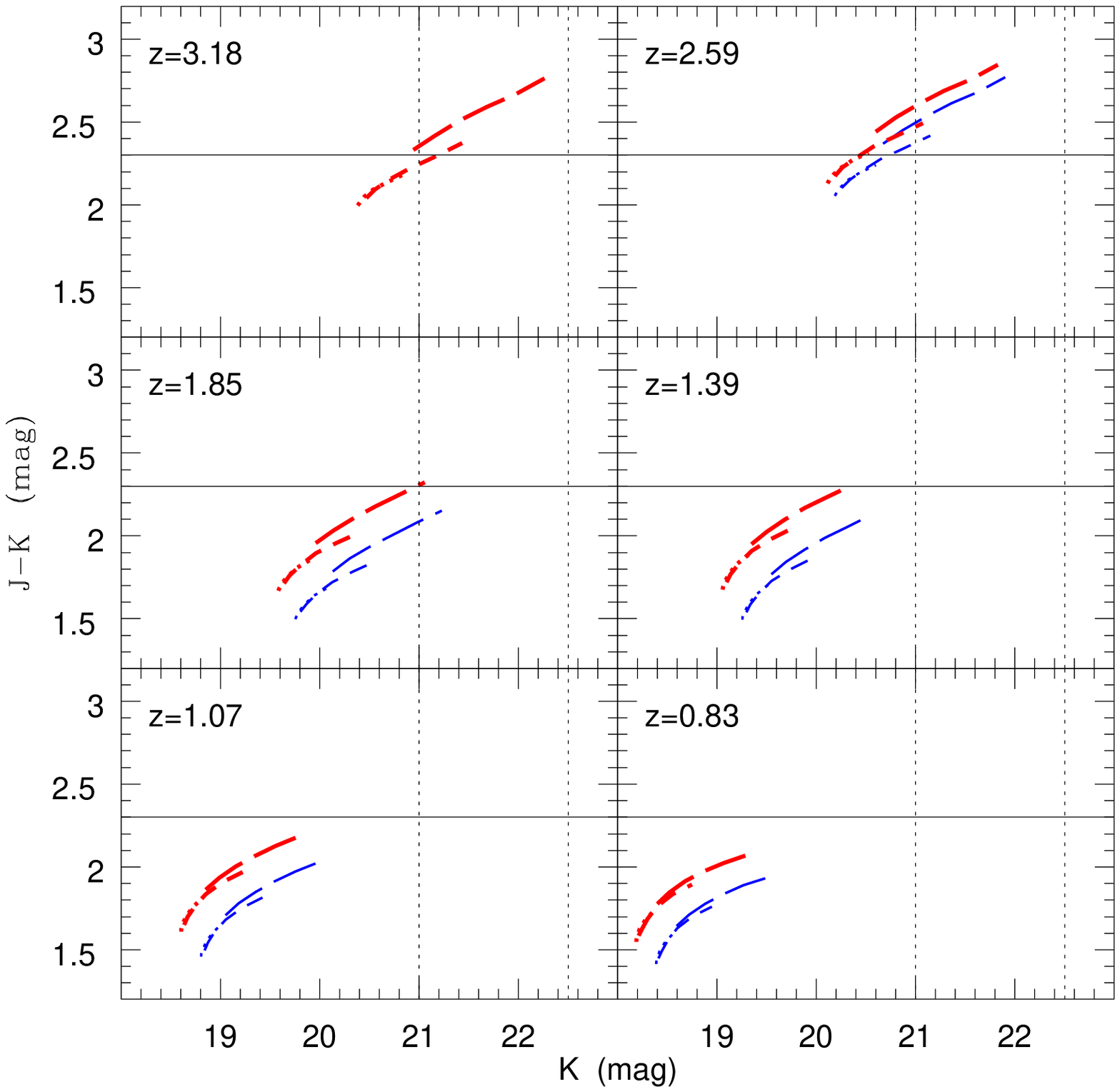}
 \caption{$\rm I_c - K$ (top) and $\rm J - K$ (bottom) vs. K magnitude
  are shown as a function of opacity, inclination, and redshift for models A1
  (red/thick lines) as an observability check. For $z \le 2.59$,
  these colour--magnitude diagrams contain also equivalent models
  in which the TP-AGB phase was neglected on purpose (blue/thin lines)
  in order to visualize the effect of the neglection of this phase.
  The stellar mass of all models is set equal to $\rm 10^{11}~M_{\sun}$
  whatever $z$.
  Dotted, short-dashed, and long-dashed lines connect models with $i = 0$\degr,
  70\degr, and 90\degr, respectively; for each line, the opacity of the model
  increases from bottom left to top right.
  In each panel, the horizontal, black solid line shows
  the appropriate colour-selection criterion,
  while the two vertical, black dotted lines
  show typical limits of present, deep K-band surveys.}
 \label{fig8}
\end{figure}

Here we discuss the observability of the models.
Figure 8 reproduces the $\rm I_c - K$--K and $\rm J - K$--K colour--magnitude
diagrams for models A1.
In addition, for $z \le 2.59$, it reproduces analogous models
in which the TP-AGB phase was neglected on purpose,
in order to visualize the effect of the neglection of this phase.
In fact, at those redshifts, the observed-frame near-IR broad-band filters
sample a portion of the intrinsic spectrum that is affected
by the TP-AGB phase.
The total stellar mass of all models is set equal to $\rm 10^{11}~M_{\sun}$
whatever $z$.
Note that the observed K luminosity at redshifts $\sim 1$ would be
25 per cent fainter if the TP-AGB phase had not been included.

Now, assuming that models A1 picture fairly well real dusty,
continuously star-forming disk-like galaxies observed at high $z$,
what should one expect for the co-moving space density of real DRdGs and ERdGs?

As Fig. 8 shows, models A1 with $\rm J - K > 2.3$
and a stellar mass of $\rm 10^{11}~M_{\sun}$ at $2 \la z \la 3.2$
meet the faintest detection limit of present deep K-band surveys
(i.e. $\sim$ 22.5 K-mag).
In particular, a few of them do so already for a K-band limit of 21 mag,
for $z \sim 2.6$.
From Fig. 8, it is straightforward to conclude that
synthetic, dusty star-forming disks like models A1 can be detected
down to 22.5 K-mag if they are more massive than $\rm \sim 10^{10}~M_{\sun}$.
Hence, if massive, bulge-less disk galaxies exist at high $z$,
and if they have properties like our models, then they contribute
up to 70 per cent (cf. Sect. 3.2.1) of the estimated surface density
of distant red galaxies ($\rm 3 \pm 0.8~arcmin^{-2}$, see Franx et al. 2003).
Furthermore, a noticeable field-to-field variation may be expected
in their counts if they are selected owing to inclination effects.
F\"orster Schreiber et al. (2004) report field-to-field variations
(by a factor of 2 on average) of the surface density
of their $\rm J - K$-selected galaxies.
However, a complete morphological information is not available
for these galaxies.
F\"orster Schreiber et al. note that cosmic variance could easily account
for a large part of the differences in the surface densities.
Conversely, if real DRdGs at $2 \la z \la 3.5$ are less massive
than $\rm 10^{10}~M_{\sun}$ (stellar mass), and if they have properties
like our models, then they would escape detection
even in the present, deepest K-band surveys.
In this case, their co-moving space-density can not be constrained
by present observations.

On the other hand, edge-on, dusty, continuously star-forming disk-like
galaxies at $1 \la z \la 2$ with $\rm I_c - K > 4$ are expected to be detected
already down to $\rm K = 21$, if their typical stellar mass
is $\rm \sim 10^{11}~M_{\sun}$ (see Fig. 8).
Furthermore, the present, deepest K-band surveys can detect analogous systems
with stellar masses as low as $\rm \sim 10^{10}~M_{\sun}$.
Thus, present $\rm I_c - K$-selected samples should provide a fair estimate
of the co-moving space-density of these systems.  

Furthermore, Fig. 8 shows that our models can explain the large fraction
of $\rm F814W - K_s$-selected, edge-on disky systems
in the Yan \& Thompson (2003) sample (with $\rm K_s < 19.5$)
only if they have stellar masses larger than $\rm \sim 10^{11}~M_{\sun}$,
and, possibly, if their characteristic star-formation time scales
are as small as 2--3 Gyr (especially for objects at $z \sim 1$).

Finally, edge-on, dusty star-forming disks at $1 \la z \la 2$
with $\rm R_c - K > 5.3$ are expected to be extremely rare (see Fig. 4 and 5),
independent of their mass (not shown).
Thus, $\rm R_c - K$-selected samples are not expected to contain
a significant fraction of dusty, star-forming, bulge-less disk galaxies
at $1 \la z \la 2$.
This is consistent with the recent observational results
of Sawicki et al. (2005).

\subsubsection{On the star-formation history of the models and the presence of a bulge in real high-$z$ red galaxies}

To reproduce the colours/SEDs of continuously star-forming galaxies
at high $z$, other authors (e.g. F\"orster Schreiber et al. 2004)
assume a constant SFR.
The $SFR(t)$ of Eq. 1 leads to a constant SFR
when $\tau_{\star}$ tends to infinity.
From the previous results we conclude that
synthetic, dusty star-forming disks with constant SFR
would not meet the colour-selection criteria for high-$z$ red galaxies,
whatever the redshift of the model and the attenuation function.
The application of Eq. 2 worsens the problem, since it implies
younger luminosity-weighted ages, hence bluer observed colours.
In conclusion the phase of maximum star-formation activity must be possibly
as short as 1 Gyr and must take place at early times.
In this case, disks possibly undergo a central gas$+$stars instability
that leads to the formation of a bulge within the same time scale however
(see Immeli et al. 2004).

As often discussed in the literature, an alternative explanation
for the red colours of real, star-forming disk galaxies at high $z$
is that they are not pure disks, but do contain a bulge.
Besides usually containing older stellar populations,
a bulge suffers from a larger dust attenuation
at rest-frame optical/near-IR wavelengths than a disk,
for intermediate/low inclinations.
The latter property is a consequence of the difference
in dust/stars configuration for the two components
(see Ferrara et at. 1999; Tuffs et al. 2004; Pierini et al. 2004b).
However before relying on this solution, one has possibly to check
the morphology.
For instance, only a small fraction of the 24 $\rm F814W - K_s$-selected
objects of the Yan et al. (2004) sample seems to be bulge-dominated.
On the other hand, of the 24 $\rm R - K_s$-selected objects with $\rm K_s < 19$
visually classified in both the R and $\rm K_s$ images simultaneously
by Sawicki et al. (2005), eleven show clear evidence for the presence of disks,
and 10 of the latter eleven\footnote{The only disk system classified
as bulge-less shows signs of interaction.} exhibit a bulge.
Finally, there is an indirect evidence for the presence of a bulge
in only 4 distant red galaxies at $z \ga 2$ (photometric, Rubin et al. 2004;
Toft et al. 2005).

\subsubsection{Geometrically-thin dust lanes at high redshift}

When the youngest stars are assumed to be the most embedded
in geometrically-thin lanes with e.g. MW-type dust (models A1 through B2),
edge-on disk models can exhibit both $\rm I_c - K > 4$ and $\rm J - K > 2.3$
for $2 \times \tau_{\rmn V} \ga 2$ and $2 \la z \la 3.2$,
or even $\rm R_c - K > 5.3$ for $2 \times \tau_{\rmn V} \ga 8$
and $1 \la z \la 3.2$ (see Fig. 4).
On the other hand, models B4, B5, B7, and B8 can exhibit $\rm I_c - K > 4$
colours only for $1 \la z < 2$, whatever $\tau_{\rmn V}$, and, eventually,
$\rm R_c - K > 5.3$ colours only for $2 \times \tau_{\rmn V} > 10$
and $z \sim 1$ (see Fig. 5 for models B5 and B8).
Conversely, models B3 and B6 never meet either the $\rm I_c - K > 4$
or the $\rm R_c - K > 5.3$ colour-selection criterion (not shown),
though they also reproduce geometrically-thin dust lanes.

Hence, the existence of edge-on, star-forming, bulge-less disk galaxies
with $\rm I_c - K > 4$ and $\rm J - K > 2.3$ at $2 < z \la 3.2$ implies that
the youngest stars are the most embedded in geometrically-thin dust lanes.

In the local Universe, disk galaxies with geometrically-thin dust lanes
seem to have rotational velocities higher than $\rm \sim 120~km~s^{-1}$
(Dalcanton et al. 2004).
Conversely, nearby disk galaxies with lower rotational velocities exhibit
a dust/stars configuration where the dust-to-stars scale-height ratio
is close to 1, i.e. as for the S01\_ME10 and S01\_SE10 dust/stars
configurations of Ferrara et al. (1999).
Furthermore, Dalcanton et al. (2004) conclude that narrower dust lanes
are due to lower characteristic turbulent velocities, as an effect
of gravitational instabilities setting in a gas$+$stars disk.
As a consequence, the disk is prone to fragmentation and gravitational collapse
along spiral arms.  

If these results apply to high-$z$ disk-like galaxies
with $\rm I_c - K > 4$ and $\rm J - K > 2.3$, these objects can be expected
to experience a relatively efficient, rapid star-formation activity.

\subsubsection{On the near-IR colours of Lyman break galaxies}

Lyman break galaxies (LBGs) at $2 \la z \la 3.4$
exhibit $\rm 0.6 < J - K < 3.2$ (Sawicki \& Yee 1998).
Their optical and near-IR colours are reproduced fairly well
by the starburst models of Vijh, Witt, \& Gordon (2003),
and seem to be due to reddening by SMC-type dust.

However, our models of dusty, continuously star-forming disks
with ages of 1.5--2 Gyr at $2 \la z \la 3.2$ exhibit $\rm 2 < J - K < 3$,
for a wide region of the explored parameter space (cf. Fig. 4 and 5).
They exhibit bluer near-IR colours (down to $\rm J - K \sim 1.6$),
if their stellar populations have maximum ages of about 0.6 Gyr (see Fig. 7).
This suggests that some of the LBGs with the reddest near-IR colours
at $2 \la z \la 3.4$ can be dusty star-forming disks
with a prolonged star-formation activity
(i.e. with intermediate-age/old stellar populations)
and not dusty young starbursts.

An accurate determination of the dynamical mass or, in its absence,
of the stellar mass for LBGs at $2 \la z \la 3.4$ can tell if the reddest ones
are relatively low-mass starburst systems or massive systems
still in the making.
An alternative discriminating tool is mid-IR (observed frame) spectroscopy,
since the rest-frame visual--near-IR SED of an intermediate-age SSP
is characterized by molecular absorption bands originated in the atmospheres
of TP-AGB stars (Maraston 2005 and references therein).

\subsubsection{On the near-IR colours of submm-sources}

Sub-mm galaxies (SMGs, or `` SCUBA sources'') with confirmed counterparts
(based on accurate positions from radio, CO, and/or millimeter continuum
interferometric observations) exhibit a median $\rm J - K = 2.6 \pm 0.6$
(Frayer et al. 2004).
In particular, the near-IR-bright (i.e. $\rm K < 19$) SMGs
exhibit $\rm J - K \simeq 2$, while the near-IR-faint SMGs
tend to have $\rm J - K > 3$.
Sub-mm sources lie at $2 \la z \la 3$ (Chapman et al. 2003a).

The unresolved, brightest sub-mm sources are held to be systems
experiencing a strong tidal interaction/merging (Chapman et al. 2003b;
Frayer et al. 2004).
However, the SMG investigated by Greve, Ivison, \& Papadopoulos (2003)
seems to be a very large (several tens of kiloparsecs) system
where the dust distribution is more extended than the stellar one,
maybe owing to the presence of strong galactic winds.
This dust/stars configuration is reminiscent of the Ferrara et al. (1999)
configuration S01\_ME25 (or S01\_SE25) that is included in models B5
(or B8, see Fig. 5).
This analogy is intriguing.
In fact, Kaviani, Haehnelt, \& Kauffmann (2003) show that models
of SCUBA sources in which the bulk of the sub-mm emission comes
from extended, cold ($\rm T \sim 20$--25 K) dust in objects
with SFRs of 50--100 $\rm M_{\sun}~yr^{-1}$ reproduce the sub-mm counts.

Therefore, it is important to realize that, according to our models,
DRdGs may have near-IR colours as red as those of some SMGs,
without being necessarily much dustier than the Milky Way.

However, none of our models achieves the reddest near-IR colours
observed so far ($\rm J - K \ga 3$--4, Maihara et al. 2001).
At least the reddest among these hyper extremely red objects
are probably primordial elliptical galaxies reddened by dust
and still in the starburst phase of their formation at $z \sim 3$
and, thus, the counterparts of the brightest sub-mm sources,
as suggested by Totani et al. (2001).

\subsubsection{On the use of single-colour selection criteria}

Figures 4 and 5 show that the overlap between $\rm R_c - K$-selected,
$\rm I_c - K$-selected, and $\rm J - K$-selected sub-samples
of synthetic, dusty star-forming disks is poor.
$\rm J - K$-selected models lie mostly at $2 \la z \la 3.2$,
while $\rm R_c - K$ and $\rm I_c - K$-selected models lie mostly
at $1 \la z \la 2$.

The adoption of arbitrary, single-colour selection criteria
introduces a bias not only in redshift but also in the intrinsic properties
of the selected objects, as intended at least in part
(see Yan \& Thompson 2003; Pierini et al. 2004a; Daddi et al. 2004;
Yan et al. 2004).
Theoretically, this bias arises from the combination
of the different rest-frame wavelength ranges probed
by the individual broad-band filters with the intrinsic properties
(stellar populations and dusty ISM) of the high-$z$ systems.

We illustrate these combined effects through the following example
based on our models of dusty, continuously star-forming disks
at $0.8 \la z \la 3.2$.

Down to a fixed K-magnitude limit, the $\rm R_c - K > 5.3$ colour-selection
criterion drops more (edge-on) disk models than the $\rm I_c - K > 4$
colour-selection criterion.
For an aging, dusty, star-forming disk at $z = 1$--2, both the $\rm R_c$
and $\rm I_c$ broad-band filters probe the stellar emission
in the rest-frame UV/U-band, dominated by OB stars, after attenuation
by internal dust.
In particular, for $z = 1$ (2), the $\rm R_c$ band maps
the rest-frame wavelength range centred at $\rm 3235~\AA$ ($\rm 2157~\AA$),
while the $\rm I_c$ band maps the rest-frame wavelength range centred
at $\rm 3932~\AA$ ($\rm 2622~\AA$).
For $z = 2$, the maximum age of the model stellar populations is about 2.8 Gyr,
and, at the systematically shorter, rest-frame near-UV wavelengths
probed by the $\rm R_c$ filter with respect to the $\rm I_c$ filter,
the increase in stellar emission overcomes the increase in dust attenuation,
whatever the dusty disk model.
In fact, the difference in attenuation at the two neighboring rest-frame
wavelengths probed by the $\rm R_c$ and $\rm I_c$ filters diminishes
when either the inclination or the opacity increases (see Ferrara et al. 1999
and Pierini et al. 2004b).
This is so even if the extinction law is of MW type,
i.e. has a local absorption peak at $\rm 2175~\AA$.
As a result, an aging, dusty star-forming disk at $z \sim 2$ turns out to be
``too blue'' for the $\rm R_c - K > 5.3$ colour-selection criterion.

On the other hand, for $z = 1$, the age of the model is about 5.3 Gyr
and the exact value of the SFR e-folding time becomes important,
since now the $\rm R_c$ filter probes also rest-frame near-UV wavelengths,
while the $\rm I_c$ filter probes rest-frame wavelengths
around the $\rm 4000~\AA$-break.
The previous considerations on dust attenuation still apply however.
As a result, an aging, dusty, star-forming disk at $z \sim 1$ turns out to be
``red enough'' for the $\rm I_c - K > 4$ colour-selection criterion.
This updates the conclusion of Yan et al. (2004) that the $\rm I_c - K$
is very effective in picking up objects with somewhat prolonged star-formation
activity (e.g. with $\tau_{\star} \rm = 1~Gyr$).

\section{Summary and conclusions}

This study focusses on those objects discovered at high $z$
that exhibit ``red'' observed-frame optical/near-infrared (IR) colours
($\rm R_c - K > 5.3$, $\rm I_c - K > 4$, or $\rm J - K > 2.3$)
and have been morphologically classified as bulge-less disk galaxies
(e.g. Yan \& Thompson 2003; Labb\'e et al. 2003).
Disks represent the building blocks of all galaxies with a spheroidal component
if the hierarchical galaxy-formation theory holds (e.g. Kauffmann 1996).
They also represent the ancestors of bulge$+$disk galaxies at $z = 0$
for the different scenarios of the so-called ``secular evolution''
of the bulge-to-disk Hubble sequence, that explain the formation
of the bulge component via different mechanisms of disk instability
(see Combes 2004 for a recent review).

The aim of this paper is to explore under which conditions of dust attenuation
and stellar populations, models with exponentially declining star-formation
histories (like those of disks in the local Universe) and without a bulge
meet the above mentioned colour-selection criteria.
These disk models start forming stars at $z_{\rmn f} = 10$
and at redshifts between 3.2 and 0.8 have ages between 1.5--6 Gyr,
solar metallicity, and Salpeter IMF.
At variance with other stellar population models, the evolutionary synthesis
code used here (Maraston 1998, 2005) includes the contribution
of the thermally pulsating phase of an intermediate-mass (2--5 M$_{\sun}$) star
along the Asymptotic Giant Branch (i.e. the TP-AGB phase).
The TP-AGB stars are fundamental for computing correctly
the emission contribution from intermediate-age (between 0.2 and 1--2 Gyr old),
simple stellar populations longward of the rest-frame V band.

The properties of dust attenuation are described by Monte Carlo calculations
of radiative transfer of the stellar and scattered radiation through different
dusty ISM for a doubly exponential disk geometry (Ferrara et al. 1999;
Pierini et al. 2004b).
The use of different sets of physically-motivated attenuation functions
allows us to probe a large fraction of the dust-attenuation parameter space,
that is still largely unconstrained by observations of high-$z$ objects.
This parameter space is defined by variables like the astrophysical properties
of the mixture of dust grains present in the ISM (synthesized
by the extinction function), the total amount of dust (given by the opacity,
i.e. $2 \times \tau_{\rmn V}$ here), the structure of the dusty ISM,
and the dust/stars configuration.
Therefore the present study complement the others in the literature
in which the so-called ``Calzetti law'' (Calzetti et al. 2000) for starbursts
is assumed in order to describe dust attenuation in all star-forming systems
at high $z$.

As a first main result, our synthetic, dusty, star-forming disks
at $0.8 \la z \la 3.2$ can exhibit red optical/near-IR colours
because of reddening by dust, but only if they have been forming stars
for more than $\sim$ 1 Gyr.
Otherwise, one has to resort to different characteristics of dust attenuation
from those of disks in the local Universe, and/or to star-formation time scales
much shorter than 3 Gyr, and/or to the existence
of an additional bulge component.

Moreover, our disk models barely exhibit $\rm R_c - K > 5.3$
at $1 \la z \la 2$.
They do so only when these three conditions are satisfied: the youngest stars
are the most embedded in a narrow dust lane; the disk is viewed
at $\sim 90$\degr; $2 \times \tau_{\rmn V} \ga 6$.
One has to resort to star-formation time scales as short as 1 Gyr
to obtain disk models with $\rm R_c - K > 5.3$ if viewed
at intermediate/low inclinations.
Such extremely short star-formation time scales might lead
to an early gas$+$stars disk instability, and, thus, to the formation
of a central bulge component (Immeli et al. 2004).
Hence, $\rm R_c - K$-selected galaxies observed at $1 \la z \la 2$ are probably
either starbursts (Pozzetti \& Mannucci 2000) or systems with a bulge,
whether they are still forming stars or not.
This conclusion is consistent with observational results
(e.g. Cimatti et al. 2002; Wilson et al. 2004; Saracco et al. 2005;
Severgnini et al. 2005; Sawicki et al. 2005).
It is also consistent with the finding that X-ray-selected AGN can exhibit
$\rm R_c - K > 5.3$ (Szokoly et al. 2004; Mignoli et al. 2004),
if the link beetwen super massive black holes and bulges
found in the local Universe holds at high $z$ (Page et al. 2004).

On the other hand, models at $1 \la z \la 2$ can exhibit $\rm I_c - K > 4$
for an inclination close to 90\degr, even for very small values of the opacity
(i.e. for $2 \times \tau_{\rmn V} > 0.5$).
This result holds for a wide range in dust/stars configuration.
It can explain the existence of a large fraction (30 per cent)
of edge-on (i.e. seen at an inclination close to 90\degr), bulge-less
disk galaxies with $\rm K_s < 19.5$ and $\rm F814W - K_s > 4$
(Yan \& Thompson 2003), some of them at $z \sim 1$ (spectroscopic,
see Yan et al. 2004).
These results imply that $\rm F814W - K_s$-selected disk galaxies
not seen edge on at $1 \la z \la 2$ do have a bulge.
Therefore, we can explain why the large majority
of the 24 bright, $\rm F814W - K_s$-selected disk galaxies
found by Yan et al. (2004) at $\rm 0.9 < z < 1.5$ has absorption features
from old stars, and why half of the systems with absorption lines
experience recent star-formation activity,
whatever their morphological classification.

$\rm I_c - K$-selected edge-on models at $1 \la z \la 2$
need to be more massive than $\rm \sim 10^{10}~M_{\sun}$ to be brighter
than the limiting magnitude of present K-band surveys (i.e. $\rm K = 22.5$).
In particular, $\rm I_c - K$-selected models at $z \sim 1$ are brighter
than $\rm K_s = 19.5$ if they have stellar masses in excess
of $\rm \sim 10^{11}~M_{\sun}$ and opacities larger than $\sim$ 2.
Therefore, our results point to the existence of massive, star-forming
disk-like galaxies at high $z$.
This is consistent with observational results (Labb\'e et al. 2003;
Stockton et al. 2004; Conselice et al. 2004).
Large disk systems may be rare at $z > 1$ (see e.g. Papovich et al. 2005),
but their existence, due to an early, rapid evolution, may explain
why the size distribution of disk galaxies at $0.25 < z < 1.25$ is consistent
with that of disk galaxies in the local Universe, and does not show
any significant evolution within this redshift range
(Ravindranath et al. 2004).

Furthermore, our models indicate that the edge-on, extremely red
(i.e. with $\rm I_c - K > 4$ or $\rm F814W - K_s > 4$), disk-like galaxies
at $1 \la z \la 2$ represent just a small fraction of those disks
that are at least 1 Gyr-old at look-back times of 7.7--10.2 Gyr.
Disks of similar ages at the same redhifts but not seen edge on are expected
to have bluer optical/near-IR colours, and, thus, elude
the observers' interest.

The previous results are consistent with the finding that the number
of $\rm R_c - K$-selected objects is less than the number
of $\rm I_c - K$-selected objects out of the same sample
of observed high-$z$ galaxies (e.g. Smail et al. 2002).
This deficit is due to the coupling of the different rest-frame wavelength
ranges covered by the $\rm R_c$ and $\rm I_c$ broad-band filters
with the properties of the different, observed stellar systems,
in terms of stellar populations (Yan \& Thompson 2002; Yan et al. 2004)
and of dust attenuation (this study; Pierini et al. 2004a).
In general, we confirm that the $\rm I_c - K > 4$ colour-selection criterion
picks up objects with evolved stellar populations (e.g. Yan et al. 2004;
McCarthy et al. 2004).
This does not imply the absence of on-going star formation however.

Synthetic, dusty, continuously star-forming disks at $2 \la z \la 3.2$
can exhibit $\rm J - K > 2.3$ without being necessarily dustier
than nearby disk galaxies (with $0.5 \la 2 \times \tau_{\rmn V} \la 2$,
Kuchinski et al. 1998).
The use of the new stellar populations evolutionary synthesis models
including the TP-AGB phase (Maraston 2005) allows to produce disks
older than 1 Gyr at $2 \la z \la 3.2$ with $\rm J - K > 2.3$, for a wide range
in inclination, dust/stars configuration, and properties of the dusty ISM.
Otherwise, one has to resort to values of $\tau_{\star}$ less than 3 Gyr
and/or to introduce a bulge component, or to consider starburst models
(e.g. Vijh et al. 2003).

In general, modeling the SED of a high-$z$ galaxy
without including the TP-AGB phase and, in addition, with the Calzetti law
(e.g. as in the work of F\"orster Schreiber et al. 2004) may bias
the best-fit solution towards older ages and/or higher values
of the attenuation (e.g. at the rest-frame V band) and/or larger masses.
For instance, our $\rm J - K$-selected models at $z \sim 2.6$
have a minimum rest-frame V-band attenuation of only 0.2 mag
(for $i = 70$\degr) and need to be more massive
than $\rm \sim 10^{10}~M_{\sun}$ (stellar mass) to be brighter
than $\rm K = 22.5$.

Finally, we note that the winds of TP-AGB stars are fundamental contributors
to the abundance of Policyclic Aromatic Hydrocarbons (PAH) in the dusty ISM
(Gillett, Forrest \& Merrill 1973).
Thus, disk-like galaxies with $\rm J - K > 2.3$ at $z > 2$
(cf. Labb\'e et al. 2003; Stockton et al. 2004) may represent dusty objects
where the stellar winds of TP-AGB stars appear as a new site for dust formation
in addition to the ejecta of Type II supernovae
and/or of pair instability supernovae (Todini \& Ferrara 2001;
Morgan \& Edmunds 2003; Nozawa et al. 2003; Hirashita et al. 2005).

\section*{Acknowledgments}
We acknowledge an anonymous referee for her/his stimulating comments.

%

\bsp

\label{lastpage}


\begin{thebibliography}{109}
\bibitem[\protect\citeauthoryear{Abraham \& Merrifield}{2000}]{abr00}
Abraham R. G., Merrifield M. R., 2000,
AJ, 120, 2835
\bibitem[\protect\citeauthoryear{Aguirre}{1999}]{agu99}
Aguirre A., 1999,
ApJ, 525, 583
\bibitem[\protect\citeauthoryear{Barnes \& Hernquist}{1992}]{bar92}
Barnes J. E., Hernquist L., 1992,
ARA\&A, 30, 705
\bibitem[\protect\citeauthoryear{Bell}{2002}]{bel02}
Bell E. F., 2002,
ApJ, 577, 150
\bibitem[\protect\citeauthoryear{Bessel \& Brett}{1988}]{bes88}
Bessel M. S., Brett J. M., 1988,
PASP, 100, 1134
\bibitem[\protect\citeauthoryear{Bianchi}{2004}]{bia04}
Bianchi S., 2004,
in Witt A. N., Clayton G. C., Draine B. T., eds,
Astrophysics of Dust. ASP, San Francisco, p. 771
\bibitem[\protect\citeauthoryear{Bianchi \& Ferrara}{2005}]{bia05}
Bianchi S., Ferrara A., 2005,
MNRAS, 358, 379
\bibitem[\protect\citeauthoryear{Bouwens et al.}{2004}]{bou04}
Bouwens R. J., Illingworth G. D., Blakeslee J. P., Broadhurst T. J.,
Franx M., 2004,
ApJ, 611, L1
\bibitem[\protect\citeauthoryear{Bruzual \& Charlot}{2003}]{bru03}
Bruzual A. G., Charlot S., 2003,
MNRAS, 344, 1000
\bibitem[\protect\citeauthoryear{Calzetti}{2001}]{cal01}
Calzetti D., 2001,
PASP, 113, 1449
\bibitem[\protect\citeauthoryear{Calzetti et al.}{1994}]{cal94}
Calzetti D., Kinney A. L., Storchi-Bergmann T., 1994,
ApJ, 429, 582
\bibitem[\protect\citeauthoryear{Calzetti et al.}{2000}]{cal00}
Calzetti D., Armus L., Bohlin R. C., Kinney A. L., Koornneef J.,
Storchi-Bergmann T., 2000,
ApJ, 533, 682
\bibitem[\protect\citeauthoryear{Chapman et al.}{2003a}]{cha03a}
Chapman S. C., Blain A. W., Ivison R. J., Smail I. R., 2003a,
Nat, 422, 695
\bibitem[\protect\citeauthoryear{Chapman et al.}{2003b}]{cha03b}
Chapman S. C., Windhorst R., Odewahn S., Yan H., Conselice C., 2003b,
ApJ, 599, 92
\bibitem[\protect\citeauthoryear{Charlot \& Fall}{1991}]{cha91}
Charlot S., Fall S. M., 1991,
ApJ, 378, 471
\bibitem[\protect\citeauthoryear{Ciardi \& Madau}{2003}]{cia03}
Ciardi B., Madau P., 2003,
ApJ, 596, 1
\bibitem[\protect\citeauthoryear{Cimatti et al.}{2002}]{cim02}
Cimatti A., et al., 2002,
A\&A, 381, L68
\bibitem[\protect\citeauthoryear{Cimatti et al.}{2003}]{cim03}
Cimatti A., et al., 2003,
A\&A, 412, L1
\bibitem[\protect\citeauthoryear{Combes}{2004}]{com04}
Combes F., 2004,
astro-ph/0406306
\bibitem[\protect\citeauthoryear{Conselice et al.}{2004}]{con04}
Conselice C. J., et al., 2004,
ApJ, 600, L139
\bibitem[\protect\citeauthoryear{Cousins}{1978}]{cou78}
Cousins A. W. J., 1978,
MNASAA, 37, 8
\bibitem[\protect\citeauthoryear{Daddi et al.}{2004}]{dad04}
Daddi E., Cimatti A., Renzini A., Fontana A., Mignoli M., Pozzetti L.,
Tozzi P., Zamorani G., 2004,
ApJ, 617, 746
\bibitem[\protect\citeauthoryear{Dalcanton et al.}{2004}]{dal04}
Dalcanton J. J., Yoachim P., Bernstein R. A., 2004,
ApJ, 608, 189
\bibitem[\protect\citeauthoryear{de Bernardis et al.}{2000}]{deb00}
de Bernardis P., et al., 2000,
Nat, 404, 955
\bibitem[\protect\citeauthoryear{Dehnen \& Binney}{1998}]{deh98}
Dehnen W., Binney J., 1998,
MNRAS, 294, 429
\bibitem[\protect\citeauthoryear{Dessauges-Zavadsky et al.}{2004}]{des04}
Dessauges-Zavadsky M., Calura F., Prochaska J. X., D'Odorico S.,
Matteucci F., 2004,
A\&A, 416, 79
\bibitem[\protect\citeauthoryear{Fall \& Efstathiou}{1980}]{fal80}
Fall S. M., Efstathiou G., 1980,
MNRAS, 193, 189
\bibitem[\protect\citeauthoryear{Ferguson et al.}{2004}]{fer04}
Ferguson H. C., et al., 2004,
ApJ, 600, L107
\bibitem[\protect\citeauthoryear{Ferrara et al.}{1999}]{fer99}
Ferrara A., Bianchi S., Cimatti A., Giovanardi C., 1999,
ApJS, 123, 437
\bibitem[\protect\citeauthoryear{F\"orster Schreiber et al.}{2004}]{for04}
F\"orster Schreiber N. M., et al., 2004,
ApJ, 616, 40
\bibitem[\protect\citeauthoryear{Franx et al.}{2003}]{fra03}
Franx M., et al., 2003,
ApJ, 587, L79
\bibitem[\protect\citeauthoryear{Frayer et al.}{2004}]{fra04}
Frayer T. D., Reddy N. A., Armus L., Blain A. W., Scoville N. Z.,
Smail I., 2004,
ApJ, 127, 728
\bibitem[\protect\citeauthoryear{Gilbank et al.}{2003}]{gil03}
Gilbank D. G., Smail I., Ivison R. J., Packham C., 2003,
MNRAS, 346, 1125
\bibitem[\protect\citeauthoryear{Gillett et al.}{1973}]{gil73}
Gillett F. C., Forrest W. J., Merrill K. M., 1973,
ApJ, 183, 87
\bibitem[\protect\citeauthoryear{Goldader et al.}{2002}]{gol02}
Goldader J. D., Meurer G., Heckman T. M., Seibert M., Sanders D. B.,
Calzetti D., Steidel C. C., 2002,
ApJ, 568, 651
\bibitem[\protect\citeauthoryear{Gordon et al.}{1997}]{gor97}
Gordon K. D., Calzetti D., Witt A. N., 1997,
ApJ, 487, 625
\bibitem[\protect\citeauthoryear{Gordon et al.}{2001}]{gor01}
Gordon K. D., Misselt K. A., Witt A. N., Clayton G. C., 2001,
ApJ, 551, 277
\bibitem[\protect\citeauthoryear{Gordon et al.}{2003}]{gor03}
Gordon K. D., Clayton G. C., Misselt K. A., Landolt A. U., Wolff M. J., 2003,
ApJ, 594, 279
\bibitem[\protect\citeauthoryear{Gordon et al.}{2004}]{gor04}
Gordon K. D. et al., 2004,
ApJS, 154, 215
\bibitem[\protect\citeauthoryear{Greve et al.}{2003}]{gre03}
Greve T. R., Ivison R. J., Papadopoulos P. P., 2003,
ApJ, 599, 839
\bibitem[\protect\citeauthoryear{Haardt \& Madau}{1996}]{haa96}
Haardt F., Madau P., 1996,
ApJ, 461, 20
\bibitem[\protect\citeauthoryear{Hernquist \& Springel}{2003}]{her03}
Hernquist L., Springel V., 2003,
MNRAS, 341, 1253
\bibitem[\protect\citeauthoryear{Hirashita et al. 2005}{2005}]{hir05}
Hirashita H., Nozawa T., Kozasa T., Ishii T. T., Takeuchi T. T., 2005,
MNRAS, 357, 1077
\bibitem[\protect\citeauthoryear{Immeli et al.}{2004}]{imm04}
Immeli A., Samland M., Gerhard O., Westera P., 2004,
A\&A, 413, 547
\bibitem[\protect\citeauthoryear{Inoue \& Kamaya}{2004}]{ino04}
Inoue A. K., Kamaya H., 2004,
MNRAS, 350, 729
\bibitem[\protect\citeauthoryear{Kajisawa \& Yamada}{2001}]{kaj01}
Kajisawa M., Yamada T., 2001,
PASJ, 53, 833
\bibitem[\protect\citeauthoryear{Kauffmann}{1996}]{kau96}
Kauffmann G., 1996,
MNRAS, 281, 487
\bibitem[\protect\citeauthoryear{Kaviani et al.}{2003}]{kav03}
Kaviani A., Haehnelt M. G., Kauffmann G., 2003,
MNRAS, 340, 739
\bibitem[\protect\citeauthoryear{Keel \& White}{2001}]{kee01}
Keel W. C., White R. E., 2001,
AJ, 122, 1369
\bibitem[\protect\citeauthoryear{Kong et al.}{2004}]{kon04}
Kong X., Charlot S., Brinchmann J., Fall S. M., 2004,
MNRAS, 349, 769
\bibitem[\protect\citeauthoryear{Kuchinski et al.}{1998}]{kuc98}
Kuchinski L. E., Terndrup D. M., Gordon K. D., Witt A. N., 1998,
AJ, 115, 1438
\bibitem[\protect\citeauthoryear{Labb\'e et al.}{2003}]{lab03}
Labb\'e I., et al., 2003,
ApJ, 591, L95
\bibitem[\protect\citeauthoryear{Madau}{1995}]{mad95}
Madau P., 1995,
ApJ, 441, 18
\bibitem[\protect\citeauthoryear{Maihara et al.}{2001}]{mai01}
Maihara T., et al., 2001,
PASJ, 53, 25
\bibitem[\protect\citeauthoryear{Mao et al.}{1998}]{mao98}
Mao S., Mo H. J., White S. D. M., 1998,
MNRAS, 297, L71
\bibitem[\protect\citeauthoryear{Maraston}{1998}]{mar98}
Maraston C., 1998,
MNRAS, 300, 872
\bibitem[\protect\citeauthoryear{Maraston}{2005}]{mar05}
Maraston C., 2005,
MNRAS, in press (astro-ph/0410207)
\bibitem[\protect\citeauthoryear{McCarthy}{2004}]{mcc04a}
McCarthy P. J., 2004,
ARA\&A, 42, 477
\bibitem[\protect\citeauthoryear{McCarthy et al.}{2004}]{mcc04}
McCarthy P. J., et al., 2004,
ApJ, 614, L9
\bibitem[\protect\citeauthoryear{Mignoli et al.}{2004}]{mig04}
Mignoli M., et al., 2004,
A\&A, 418, 827
\bibitem[\protect\citeauthoryear{Mo et al.}{1998}]{mo98}
Mo H. J., Mao S., White S. D. M., 1998,
MNRAS, 295, 319
\bibitem[\protect\citeauthoryear{Moll\'a et al.}{1997}]{mol97}
Moll\'a M., Ferrini F., \& D\'iaz A. I., 1997,
ApJ, 475, 519
\bibitem[\protect\citeauthoryear{Moll\'a et al.}{2000}]{mol00}
Moll\'a M., Ferrini F., Gozzi G.,  2000,
MNRAS, 316, 345
\bibitem[\protect\citeauthoryear{Morgan \& Edmunds}{2003}]{mor03}
Morgan H. L., Edmunds M. G., 2003,
MNRAS, 343, 427
\bibitem[\protect\citeauthoryear{Moustakas et al.}{2004}]{mou04}
Moustakas L. A., et al., 2004,
ApJ, 600, L131
\bibitem[\protect\citeauthoryear{Nozawa et al.}{2003}]{noz03}
Nozawa T., Kozasa T., Umeda H., Maeda K., Nomoto K., 2003,
ApJ, 598, 785
\bibitem[\protect\citeauthoryear{Page et al.}{2004}]{pag04}
Page M. J., Stevens J. A., Ivison R. J., Carrera F. J., 2004,
ApJ, 611, L85
\bibitem[\protect\citeauthoryear{Papovich et al.}{2005}]{pap05}
Papovich C., Dickinson M., Giavalisco M., Conselice C. J.,
Ferguson H. C., 2005,
astro-ph/0501088
\bibitem[\protect\citeauthoryear{Persson et al.}{1983}]{per83}
Persson S. E., Aaronson M., Cohen J. G., Frogel J. A., Matthews K., 1983,
ApJ, 266, 105
\bibitem[\protect\citeauthoryear{Pierini et al.}{2004a}]{pie04a}
Pierini D., Maraston C., Bender R., Witt A. N., 2004a,
MNRAS, 347, 1
\bibitem[\protect\citeauthoryear{Pierini et al.}{2004b}]{pie04b}
Pierini D., Gordon K. D., Witt A. N., Madsen G. J., 2004b,
ApJ, 617, 1022
\bibitem[\protect\citeauthoryear{Pozzetti \& Mannucci}{2000}]{poz00}
Pozzetti L., Mannucci F., 2000,
MNRAS, 317, L17
\bibitem[\protect\citeauthoryear{Ravindranath et al.}{2004}]{rav04}
Ravindranath S. et al., 2004,
ApJ, 604, L9
\bibitem[\protect\citeauthoryear{Renzini}{2004}]{ren04}
Renzini A., 2004,
astro-ph/0410295
\bibitem[\protect\citeauthoryear{Rubin et al.}{2004}]{rub04}
Rubin K. H. R., van Dokkum P. G., Coppi P., Johnson O.,
F\"orster Schreiber N. M., Franx M., van der Werf P., 2004,
ApJ, 613, L5
\bibitem[\protect\citeauthoryear{Salpeter}{1955}]{sal55}
Salpeter E., 1955,
ApJ, 121, 161
\bibitem[\protect\citeauthoryear{Saracco et al.}{2005}]{sar05}
Saracco P., et al., 2005,
MNRAS, 357, L40
\bibitem[\protect\citeauthoryear{Sawicki \& Yee}{1998}]{saw98}
Sawicki M., Yee H. K. C., 1998,
AJ, 115, 1329
\bibitem[\protect\citeauthoryear{Sawicki et al.}{2005}]{saw05}
Sawicki M., Stevenson M., Barrientos L. F., Gladman B., Mall\'en-Ornelas G.,
van den Bergh S., 2005,
astro-ph/0503523
\bibitem[\protect\citeauthoryear{Schneider et al.}{2002}]{sch02}
Schneider R., Ferrara A., Natarajan P., Omukai K., 2002,
ApJ, 571, 30
\bibitem[\protect\citeauthoryear{Severgnini et al.}{2005}]{sev05}
Severgnini P., et al., 2005,
A\&A, 431, 87
\bibitem[\protect\citeauthoryear{Shapley et al.}{2004}]{sha04}
Shapley A. E., Erb D. K., Pettini M., Steidel C. C., Adelberger K. L., 2004,
ApJ, 607, 226
\bibitem[\protect\citeauthoryear{Silva et al.}{1998}]{sil98}
Silva L., Granato G. L., Bressan A., Danese L., 1998,
ApJ, 509, 103
\bibitem[\protect\citeauthoryear{Smail et al.}{2002}]{sma02}
Smail I., Owen F. N., Morrison G. E., Keel W.C., Ivison R. J.,
Ledlow M. J., 2002,
ApJ, 581, 844
\bibitem[\protect\citeauthoryear{Spergel et al.}{2003}]{spe03}
Spergel D. N., et al., 2003,
ApJS, 148, 175
\bibitem[\protect\citeauthoryear{Stevens et al.}{2003}]{stev03}
Stevens J. A., Page M. J., Ivison R. J., Smail I., Lehmann I., Hasinger G.,
Szokoly G., 2003,
MNRAS, 342, 249
\bibitem[\protect\citeauthoryear{Stockton et al.}{2004}]{sto04}
Stockton A., Canalizo G., \& Maihara T., 2004,
ApJ, 605, 37
\bibitem[\protect\citeauthoryear{Szokoly et al.}{2003}]{szo03}
Szokoly G. P., et al., 2004,
ApJS, 155, 271
\bibitem[\protect\citeauthoryear{Tinsley}{1980}]{tin80}
Tinsley B. M., 1980,
Fund. Cos. Phys., 11, 1
\bibitem[\protect\citeauthoryear{Todini \& Ferrara}{2001}]{tod01}
Todini P., Ferrara A., 2001,
MNRAS, 325, 726
\bibitem[\protect\citeauthoryear{Toft et al.}{2005}]{tof05}
Toft S., van Dokkum P. G., Franx M., Thompson R. I., Illingworth G. D.,
Bouwens R. J., Kriek M., 2005,
ApJ, 624, L9
\bibitem[\protect\citeauthoryear{Totani et al.}{2001}]{tot01}
Totani T., Yoshii Y., Iwamuro F., Maihara T., Motohara K., 2001,
ApJ, 558, L87
\bibitem[\protect\citeauthoryear{Tuffs et al.}{2004}]{tuf04}
Tuffs R. J., Popescu C. C., V\"olk H. J., Kylafis N. D., Dopita M. A., 2004,
A\&A, 419, 821
\bibitem[\protect\citeauthoryear{V\"ais\"anen \& Johansson}{2004}]{vai04}
V\"ais\"anen P., Johansson P. H., 2004,
A\&A, 421, 821
\bibitem[\protect\citeauthoryear{van Dokkum et al.}{2004}]{vdok04}
van Dokkum P. G., et al., 2004,
ApJ, 611, 703
\bibitem[\protect\citeauthoryear{Vijh et al.}{2003}]{vij03}
Vijh U. P., Witt A. N., Gordon K. D., 2003,
ApJ, 587, 533
\bibitem[\protect\citeauthoryear{White et al.}{2000}]{whi00}
White R. E., Keel W. C., Conselice C. J., 2000,
ApJ, 542, 761
\bibitem[\protect\citeauthoryear{White \& Rees}{1978}]{whi78}
White S. D. M., Rees M. J., 1978,
MNRAS, 183, 341
\bibitem[\protect\citeauthoryear{White \& Frenk}{1991}]{whi91}
White S. D. M., Frenk C. S., 1991,
ApJ, 379, 52
\bibitem[\protect\citeauthoryear{Whittet}{2003}]{whi03}
Whittet D. C. B., 2003,
Dust in the galactic environment, 2nd edn. Institute of Physics, Bristol
\bibitem[\protect\citeauthoryear{Wilson et al. 2004}{2004}]{wil04}
Wilson G., et al., 2004,
ApJS, 154, 107
\bibitem[\protect\citeauthoryear{Witt \& Gordon}{1996}]{wit96}
Witt A. N., Gordon K. D., 1996,
ApJ, 463, 681
\bibitem[\protect\citeauthoryear{Witt \& Gordon}{2000}]{wit00}
Witt A. N., Gordon K. D., 2000,
ApJ, 528, 799
\bibitem[\protect\citeauthoryear{Yan \& Thompson}{2003}]{yan03}
Yan L., Thompson D., 2003,
ApJ, 586, 765
\bibitem[\protect\citeauthoryear{Yan et al.}{2004}]{yan04}
Yan L., Thompson D., Soifer B. T., 2004,
AJ, 127, 1274
\end{thebibliography}
\end{document}